\documentclass[aps,onecolumn,preprint,prd,superscriptaddress,nofootinbib]{revtex4}
\usepackage{graphicx}
\usepackage{amsmath,amssymb,amsthm}
\usepackage{latexsym,graphicx,color,subfigure,slashed,mathrsfs,verbatim,bbm,wasysym, pstricks,epsfig,url, hyperref}
\usepackage{epstopdf}
\usepackage{nicefrac}

\newcommand{\beq}{\begin{eqnarray}}
\newcommand{\eeq}{\end{eqnarray}}

\newcommand{\cL}{{\cal L}}

\newcommand{\cH}{{\cal H}}
\newcommand{\hQ}{{\hat Q}}
\newcommand{\hu}{{\hat u}}

\newcommand{\hyt}{{\hat y_t}}
\newcommand{\hg}{{\hat g}}

\newcommand{\hL}{{\hat \Lambda_{3}}}
\newcommand{\met}{{{\slashed E_T}}}

\newcommand{\zpp}{{G_{0+}}}
\newcommand{\zmp}{{G_{0-}}}
\newcommand{\ztwo}{{\mathbb Z}_2}
\newcommand{\hb}{{\hat b}}

\def\btwin{{\hat b}}

\def\gtwin{{\hat g}}
\def\GeV{\ {\rm GeV}}
\def\Br{\ {\rm Br}}
\def\bbtwin{[\btwin\bar\btwin]}
\def\etab{{\hat\eta}}     
\def\chib{{\hat \chi}}

\begin{document}

\begin{titlepage}
 
\thispagestyle{empty}

\begin{flushright}UMD-PP-014-028\\
CERN-PH-TH-2014-263
\end{flushright}
\vspace{0.2cm}
\begin{center} 
\vskip 2.0cm
{\LARGE \bf Naturalness in the Dark at the LHC}
\vskip 0.5cm

\vskip 1.0cm
{\large Nathaniel Craig$^{1}$, Andrey Katz$^{2,3,4}$,\\ Matt Strassler$^4$, and Raman Sundrum$^5$}
\vskip 0.4cm
{\it $^1$Department of Physics, University of California, Santa Barbara, CA 93106}\\
{\it $^2$Theory Division, CERN, CH-1211 Geneva 23, Switzerland}\\
{\it $^3$Universit\'e de Gen\`eve, Department of Theoretical Physics and Center for Astroparticle Physics (CAP), 
24 quai E. Ansermet, CH-1211, Geneva 4, Switzerland}\\
{\it $^4$ Department of Physics, Harvard University, Cambridge, MA 02138}\\
{\it $^5$Department of Physics, University of Maryland, College Park, MD 20742}\\
\vskip 1.7cm
\end{center}

\noindent 
We revisit the Twin Higgs scenario as a ``dark'' solution to the little hierarchy problem, identify the structure of  a 
 minimal model and its viable parameter space, and analyze its collider implications.  In this model, dark naturalness generally leads to  Hidden Valley phenomenology.
The twin particles, including the top partner, are all Standard-Model-neutral,  but naturalness favors the existence 
of twin strong interactions -- an asymptotically-free force that confines not far above the Standard Model QCD scale -- and a Higgs portal interaction.  We show that, taken together, these typically give rise to exotic decays of the Higgs to twin hadrons.   Across a substantial portion of the parameter space, certain twin hadrons have visible and often displaced decays, providing a potentially striking LHC signature.  We briefly discuss appropriate experimental search strategies.

\end{titlepage}

\tableofcontents

\section{Introduction}
\label{sec:intro}

The principle of Naturalness, the notion that the weak scale should be insensitive to quantum effects from physics at much higher mass scales, necessitates new TeV-scale physics beyond the Standard Model (SM). It has motivated a broad program of searches at the LHC, as well as lower-energy precision/flavor/CP experiments. 
The absence thus far of any signals in these experiments has disfavored 
the most popular scenarios, including supersymmetry (SUSY), Composite Higgs and Extra Dimensions, unless their mass scales are raised above natural expectations, leading to sub-percent level electroweak fine-tuning in complete models.  The puzzle over why Nature should be tuned at this level has been dubbed the ``Little Hierarchy Problem,"  
 and it has led to a major rethinking of naturalness and its implications for experiments.

In a bottom-up approach to this problem, one may take a relatively agnostic view of very high energy physics, and focus instead on naturalness of just the ``little hierarchy",  from an experimental cutoff of about 5-10 TeV or so down to the weak scale.   Unlike naturalness considerations involving extremely high scales, such as the Planck scale, which are tied to multi-loop virtual effects on the Higgs sector of all SM particles, the bottom-up little hierarchy problem is simpler, relating predominantly to one-loop effects of just the heaviest  SM particles, {\it i.e.} those coupling most strongly to the Higgs. In Little Higgs, (Natural) SUSY, and extra-dimensional models of gauge-Higgs unification (including warped models that are dual to Higgs Compositeness via the AdS/CFT correspondence), large one-loop radiative corrections to the Higgs from the heaviest SM particles cancel algebraically against those of new symmetry ``partners" of these heavy particles, thereby ensuring stability of the little hierarchy. 
   
The most significant of these corrections is associated to the top quark.  Naturalness requires that this must be substantially canceled by a corresponding correction from the top's partner(s), to which it is related by an ordinary global symmetry or supersymmetry.  This requirement can only be fulfilled naturally if the associated partner has a mass scale $\sim 500$ GeV, easily within the kinematic reach of the LHC. Such particles also have significant LHC production cross-sections, since a top partner generally carries the same color charge as the top quark. Through this logic, the search for top partners in the above incarnations, under a variety of assumptions about possible decay modes, has become  a central pursuit of the LHC.

But {\it must} the top partner be colored? The answer is obviously critical to experimental exploration. Naively the answer is yes, because the algebraic cancellations depend at one-loop on the top Yukawa coupling to the Higgs, and this coupling is itself corrected at one higher loop by QCD. In a rare exception to the ``one-loop rule'', such two-loop radiative corrections to the Higgs are still quantitatively important for the little hierarchy problem because of the strength of QCD. It would seem then that the top partner should also be colored so as to parametrically ``know'' about this QCD effect in symmetry-enforced cancellations. 

Yet, remarkably, there do exist solutions to the little hierarchy problem, ``Twin Higgs" being the first and prime example,  in which the top partners are 
uncolored~\cite{Chacko:2005pe,Barbieri:2005ri,Chacko:2005vw}.\footnote{Another known solution to the
little hierarchy problem which involves uncolored top partners is 
``Folded Supersymmetry"~\cite{Burdman:2006tz}.
}
Here the cancellation among radiative corrections is enforced by a discrete $\ztwo$ symmetry that exchanges SM particles with new ``twin" states. One way to assure naturalness cancellation of QCD two-loop effects is to have the twin symmetry exchange SM color and its gluons with a distinct twin color gauge group and its twin gluons, which couple to and correct the twin top partner just  as QCD does the top quark.  We will focus on theories of this type, specifically ones in which all twin particles are ``dark'', with no SM quantum numbers.

Colorless twin tops are vastly more difficult to produce at the LHC than top partners of more popular theories; indeed this is true for all twin sector particles. Twin particle production can only proceed through a ``Higgs portal'', a modest mixing between the SM and twin Higgs sectors that is a necessary consequence of the twin Higgs mechanism for addressing the little hierarchy.  Not only is the production rate small, the hidden particles barely interact with ordinary matter, and (at least naively) one would expect they escape the detectors unobserved. The resulting missing energy signature with a very low cross-section  would pose great difficulties at the LHC.

How else can the twin sector be detected?  Higgs mixing and virtual twin top loops can 
also subtly affect the SM-like Higgs, making precision tests of its properties extremely important.  
But with expected LHC precision, the visibility of the twin sector in this manner at the LHC is limited~\cite{Burdman:2014zta}. The Higgs of the twin 
sector can also potentially be produced, but it may be too wide to observe as a resonance.  At best, it is heavy and has a low cross-section, and is far from being excluded or discovered.  For these reasons, Twin Higgs remains a viable resolution of the little hierarchy problem, a well-hidden outpost of naturalness.

 In this paper, we re-examine the Twin Higgs scenario as an important and distinctive case study in ``dark" naturalness, and we identify exciting new experimental opportunities for its discovery. We develop {\it minimal} Twin Higgs models addressing the little hierarchy problem, roughly paralleling the way in which  
 ``Natural SUSY''~\cite{Dimopoulos:1995mi,Cohen:1996vb,Sundrum:2009gv,Barbieri:2010pd,Papucci:2011wy,Brust:2011tb}  
 has emerged as a minimal phenomenological approach to   the little hierarchy problem in the SUSY paradigm.  In both cases, minimalism can be viewed as a tentative organizing principle for doing phenomenology, starting with searches for the minimal natural spectrum and then ``radiating outwards" to include searches complicated by non-minimal states.  Also in parallel to SUSY, we find that the two-loop relevance of QCD interactions to the little hierarchy problem leads to some of the most promising experimental signals. 

 In SUSY, the symmetry cancellation at two loops requires the presence of a gluon-partner, the gluino.  With large color charge and spin, the gluino is phenomenologically striking over much of motivated parameter  space, almost independent of its decay 
 modes~\cite{Evans:2013jna,Berger:2013sir,Han:2012cu}.  In Twin Higgs models, the analogous two-loop role is played by twin gluons, which can again  give rise to striking signatures over a large part of parameter space, not because of large cross-sections but because they, along with any light twin matter, are confined into bound states: twin hadrons. 
 Together with the Higgs portal connecting the SM and twin sectors, the presence of metastable hadrons sets up classic ``confining Hidden Valley" 
 phenomenology~\cite{Strassler:2006im,Strassler:2006qa,Strassler:2006ri,Han:2007ae,Strassler:2008bv,Strassler:2008fv,Juknevich:2009ji},
 now in a plot directly linked to naturalness.

 	\begin{figure}
	\centering
	\includegraphics[width=0.7\textwidth]{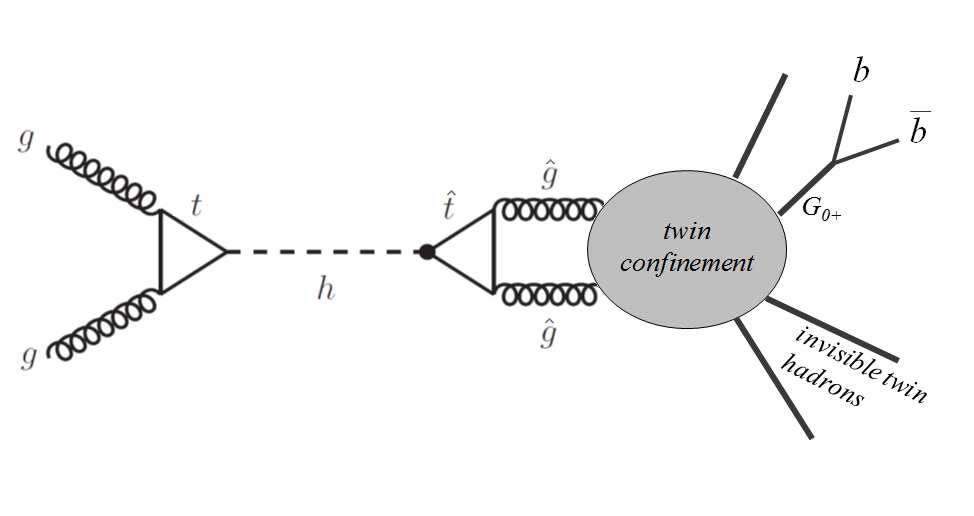}
              \caption{Example of a Twin Higgs collider event. The SM-like Higgs decays through a loop of the twin tops
       into a pair of twin gluons, which subsequently hadronize to produce various twin glueballs. While some glueballs are stable at the 
       collider scale, $\zpp$ decay to Standard Model particles is sufficiently fast to give LHC-observable effects, including possible displaced vertices. The $h\gtwin\gtwin$ coupling, indicated by a black dot, is generated by small mixing of the Higgs and the twin Higgs.}\label{fig:makingglueballs}
	\end{figure}
        
 A prototypical new physics event is illustrated in 
Fig.~\ref{fig:makingglueballs}.
The scalar line represents the recently discovered $125$~GeV Higgs 
scalar.
This particle is primarily the 
SM Higgs with a small admixture 
of twin Higgs; it is readily produced by gluon fusion.  But because of its twin Higgs content, it has at least one exotic decay mode into twin gluons, induced by twin top loops, with a branching fraction of order~0.1\%.
The twin gluons  ultimately hadronize into twin glueballs, which have mass in the 
$\sim 1 - 100$~GeV range within the minimal model. While most twin glueballs have very long lifetimes and 
escape the detector as missing energy, {\it the lightest $0^{++}$ twin glueball has the right quantum numbers 
to mix with the SM Higgs, allowing it to decay back to the SM on detector timescales.}  The first excited $0^{++}$ state also may have this property.
This type of effect was first studied, in the context of a Hidden Valley/quirk 
model~\cite{Strassler:2006im,Strassler:2006ri,Kang:2008ea,Juknevich:2009ji} by Juknevich~\cite{Juknevich:2009gg}.   

If the lightest $0^{++}$ glueball, which we call $\zpp$, has a high mass, then its decay is prompt, and its production rate in Higgs decays may be too rare for it to be observed at the LHC.  However, these non-SM decays of the Higgs would be discoverable at  $e^+e^-$ machines, providing strong motivation for ILC, FCC-ee, or CEPC.

For lower mass (typically below $\sim 40$ GeV in the minimal model we describe) the decay of the $\zpp$ may be macroscopically displaced from the interaction point. Such displaced decays are a striking 
signature, spectacular enough to compensate for the relatively low production rate, and represent an excellent opportunity for the LHC.  

Moreover, the signal may be enhanced and/or enriched if there is a sufficiently light twin bottom quark.  Depending on this quark's mass, twin bottom production may lead to a larger overall twin hadron production rate; the resulting final states may include twin glueballs, twin bottomonium, or both.
In some portions of parameter space, all twin bottomonium states are invisible.  In others the lightest $0^{++}$ state (which we will call $\chib$, since it is analogous  to the $\chi_0$ states of SM quarkonium)
can decay visibly by mixing with the Higgs just like the $\zpp$, though possibly with a small branching fraction.  There is also the potential for displaced vertices from this state, though our calculations of lifetimes and branching fractions suffer from large uncertainties.

With a branching fraction of the Higgs to twin hadrons  of order $0.1\%$ or greater,  our minimal Twin Higgs model should motivate further  experimental searches for this signal of hidden naturalness.  A branching fraction of $0.1\%$ represents $\sim 500$ events produced at ATLAS and at CMS during Run I, and we can expect many more at Run II.   But triggering inefficiencies threaten this signal.  Despite the low backgrounds for highly displaced vertices, triggering on such events can be a significant 
challenge~\cite{Strassler:2006qa,jhudisplaced,ATLASTriggerNote,ATLASLLPTriggers,ATLAShvMS2,ATLAShvHCAL,CMSLLPjets,Buckley:2014ika,Cui:2014twa}.  There is urgency as we approach Run II, since it is not trivial to design triggers that will efficiently capture  all variants of the displaced vertex signature arising within the parameter space of the model.
In particular, since  twin hadron lifetimes  are a strong function of their mass, and since displaced 
decays in different subdetectors require quite different trigger and analysis methods, a considerable variety 
of approaches will be required for efficient coverage of the parameter space of the model.  Further complicating 
the matter is that precise theoretical predictions of twin glueball and especially twin quarkonium  production in 
Higgs decays is extremely difficult.  
Hadronization in the twin sector is a complex and poorly understood process, 
and considerably more investigation will be needed before predictions of 
glueball and quarkonium multiplicity and kinematic distributions could be possible.  Thus this solution to the naturalness problem requires further work on both experimental and theoretical fronts.

We should note that every element that goes into this story has appeared previously in the literature.  Hidden glueballs appear in quirk models and other Hidden Valley 
models~\cite{Strassler:2006im,Kang:2008ea,Juknevich:2009ji}; they specifically arise in Folded Supersymmetry \cite{Burdman:2006tz} and the Quirky Little Higgs \cite{Cai:2008au}, which attempt to address the hierarchy problem; the mixing of the $\zpp$ with the Higgs to generate a lifetime that is short enough to be observed at the LHC but long enough to often be displaced has appeared in a study \cite{Juknevich:2009gg} of a Hidden Valley model with quirks \cite{Juknevich:2009ji}.   However we believe that this is the first time these elements have all been assembled together, giving the striking observable signal of exotic Higgs decays, possibly to long-lived particles, as
a sign of hidden naturalness.

Returning from the phenomenology to broader considerations, we note that the Twin Higgs shares with Little Higgs and Composite Higgs models  the realization of the Higgs as a pseudo-Goldstone boson, while representing a significant break with earlier thinking and phenomenology in having a colorless top partner. In the far UV, it is possible that the Twin Higgs structure might match on  to more conventional 
SUSY~\cite{Falkowski:2006qq,Chang:2006ra,Craig:2013fga} 
 or Composite dynamics~\cite{Batra:2008jy,Geller:2014kta}, or perhaps to something quite 
 novel~\cite{Craig:2014aea,Craig:2014roa}.  
We hope that studying the UV matching is facilitated by our bottom-up exploration of minimal Twin Higgs structure and phenomenology. We also hope that this work broadens our perspective on naturalness, motivates more careful investigations of other ``darkly" natural mechanisms, such as 
``Folded Supersymmetry"~\cite{Burdman:2006tz}, and perhaps inspires entirely new mechanisms. Ultimately, we hope to broaden the scope of experimental strategies motivated by naturalness, perhaps even leading to a discovery unanticipated by theory.

The paper is organized as follows.   Section \ref{sec:min}
reviews the basic Twin Higgs structure and develops the minimal  model. 
We show in Section \ref{sec:tuning} that the electroweak tuning in the minimal Twin Higgs model is very mild.
Section~\ref{sec:color} shows quantitatively that the twin sector likely contains twin QCD in order to maintain naturalness, and its confinement and hadrons are discussed.
Section~\ref{sec:twinhadronpheno} derives some properties of the resulting twin hadrons, and discusses their production and decays. Some subtle points on hadron decays and hadron production are left for Appendices \ref{app:quarkonium} and \ref{app:hadroproducdetails}.  Section \ref{sec:LHCpheno} synthesizes the earlier considerations and discusses LHC phenomenological implications.  Possible experimental strategies for long-lived particle searches are briefly considered in Section \ref{subsec:zppdecays} and in more detail in Appendix \ref{app:strategies}.  Precision Higgs measurements are considered in Section \ref{subsec:precisionhiggs}, and precision electroweak constraints are discussed in Appendix \ref{app:precisionEW}.  Our conclusions appear in Section \ref{sec:conclu}.  In Appendix \ref{app:hypercharge}, the phenomenological effect of gauging twin hypercharge, as a non-minimal extension of the model, is considered.


\section{The Minimal  or ``Fraternal" Twin Higgs} \label{sec:min}

In this section we construct the {\it minimal} Twin Higgs model, starting by reviewing the basic symmetries and Higgs structure and then justifying each addition to the twin sector based on the need to maintain naturalness and internal consistency.   Because the minimal model does not duplicate {\it all} SM states in the twin sector (in contrast to the original mirror Twin Higgs model and its descendants~\cite{Chacko:2005pe, Barbieri:2005ri, Chacko:2005vw}, in which the twin sector and its couplings are an exact copy of the SM), we will refer to this construction as the ``Fraternal Twin Higgs". The model is summarized in subsection
 \ref{subsec:summaryofmodel}.

\subsection{The Central Mechanism}

At its heart, the Twin Higgs  mechanism involves realizing the SM-like Higgs as a pseudo-Goldstone boson of 
an approximate  global symmetry, namely  $SU(4)$.
 An $SU(4)$-fundamental complex scalar $\mathcal{H}$ with potential, 
\beq \label{eq:simple}
V =  \lambda (|\mathcal{H}|^2 - f^2/2)^2 \,,
\eeq
acquires a vacuum expectation value $\langle \mathcal{H} \rangle = f/\sqrt{2}$, breaking $SU(4) \to SU(3)$ and giving rise to seven Goldstone bosons, of which one is ultimately identified as the SM-like Higgs scalar. The $SU(4)$ is explicitly broken by the gauge and Yukawa couplings of the Standard Model. Without additional  recourse, this explicit breaking would lead to quadratic sensitivity to higher scales at one loop.

In the context of conventional global symmetry protection, this explicit breaking and UV sensitivity can be ameliorated by extending Standard Model gauge bosons and fermions into representations of the full global symmetry, at the price of introducing additional partner states charged under the SM. In the Twin Higgs, the key insight is that SM states need {\it not} be extended into full representations of the global symmetry, and instead are merely related to partner states by a ${\mathbb Z}_2$ exchange symmetry.   This ${\mathbb Z}_2$ is then promoted, at the level of quadratically divergent radiative corrections,  to an accidental $SU(4)$ symmetry.  The partner particles are no longer related to SM states by a continuous symmetry, and so need not carry SM gauge quantum numbers.

As in any global symmetry mechanism that stabilizes the weak scale, the Twin Higgs does not address the big hierarchy problem all the way to the Planck scale, but merely a little hierarchy up to a cutoff $\Lambda$, which is typically $\sim 5-10 $~TeV. This also roughly matches the maximal reach of the LHC.
Above $\Lambda$ we imagine that one of the canonical solutions to the ``big'' hierarchy problem, such as supersymmetry or compositeness, kicks in to provide protection against yet higher scales.

We can embed the SM Higgs into (\ref{eq:simple}) by decomposing  $\cH = (A, B)$ into two doublets $A$ and $B$.
We identify $A$ with the SM Higgs doublet of gauged electroweak $SU(2)_L$  (and charged under $U(1)_Y$ of course), while $B$ is a doublet of a \emph{different} $SU(2)$ symmetry, which we will call twin $SU(2)$. At this stage, the twin $SU(2)$ group can be either global or gauged. 
The ${\mathbb Z}_2$ symmetry acts by exchanging $A \leftrightarrow B$.
By far the largest source of explicit breaking from the Standard Model will be the top Yukawa, so to see the magic of the twin mechanism let us also introduce twin top multiplets: three species of the twin fermion $\hQ^a$ transforming as doublets of the twin
$SU(2)$  as well as twin right-handed tops $\hu^a$, where $a = 1 \cdots 3$. The ${\mathbb Z}_2$ symmetry implies a twin top Yukawa coupling
\beq\label{eq:twintopcouple}
\cL \supset \hyt B \hQ^a \hu^a \,, 
\eeq  
where we expect $\hyt \sim y_t$ for the twin mechanism to be effective. Note that the $a$ index implies at least a global $SU(3)$ symmetry acting on fermions $\hQ$ and $\hu$.

The one-loop radiative potential for the $A$ and $B$ multiplets coming from loops of top and twin top quarks takes the form
\beq \label{eq:oneloop}
16 \pi^2 V^{(1-loop)} = - 6 y_t^2 \Lambda^2 |A|^2 -   6 \hyt^2 \Lambda^2 |B|^2 + 3 y_t^4 |A|^4 \log \left(\nicefrac{\Lambda^2}{ y_t^2 |A|^2} \right) + 3 \hyt^4 |B|^4 \log \left( \nicefrac{\Lambda^2}{\hyt^2 |B|^2} \right)
\eeq
for a uniform cutoff $\Lambda$.\footnote{ One should interpret this cutoff as merely a proxy for physical effects;  in a realistic UV completion $\Lambda$ will be replaced by $\mathbb{Z}_2$-symmetric physical thresholds.} Notice that if $y_t = \hyt$ then the terms quadratic in the cutoff arrange themselves into an $SU(4)$ invariant $\propto |\cH|^2$, while the radiative quartics explicitly break the $SU(4)$ but preserve the $\mathbb{Z}_2$. This is the magic of the twin mechanism: if the SM-like Higgs can be identified with a pseudo-Goldstone of the spontaneously broken $SU(4)$, its mass will be insensitive to the $SU(4)$-symmetric quadratic divergences 
at one loop (or more) provided the $\mathbb{Z}_2$ relates $y_t$ and $\hyt$.

A second central issue is vacuum alignment. The vacuum expectation value (vev) of 
$\cH$, $f/\sqrt{2}$,  which breaks the approximate global $SU(4)$ symmetry down to $SU(3)$, can be decomposed,
\begin{equation}
f^2 = v_A^2 + v_B^2\ . 
\end{equation}
In general this vev breaks both $SU(2)_L \times U(1)_Y$ and twin $SU(2)$, such that most components of $\cH$ are eaten if twin $SU(2)$ is gauged. The remaining physical scalar states consist of linear combinations of the radial mode and a single uneaten Goldstone boson. We will identify the observed SM-like  Higgs with the uneaten Goldstone boson 
of $SU(4)/SU(3)$. The mass 
of the radial mode (corresponding to $|{\cal H}|^2$ fluctuations) is $\sqrt{2 \lambda} f$. 
(Alternately, we could work purely in terms of the non-linear sigma model of $SU(4)/SU(3)$, in which case there is no perturbative radial mode, and $\Lambda$ is identified with the cutoff of the non-renormalizable theory. We do not follow this approach here.)
To obtain a realistic vacuum, we must break ${\mathbb Z}_2$ to a small extent. 
In an exact ${\mathbb Z}_2$ model the potential favors $v_A = v_B$ and the pseudo-Goldstone Higgs is an equal mixture of $A$ and $B$, where recall only $A$ carries SM Higgs quantum numbers. 
 Since we have observed a SM-like Higgs experimentally, we will need perturbations stabilizing $v_A^2 \ll v_B^2$, so that the 
 pseudo-Goldstone Higgs is primarily aligned with $A$.

\subsection{Minimal Particle Content}

Thus far we have seen how the essential structure of the twin mechanism operates at the level of the Higgs and the largest source of explicit $SU(4)$ breaking, the top Yukawa coupling, and understood the new twin states thereby required. We continue this process of deducing the minimal ingredients for a realistic and natural Twin Higgs model. 

\begin{figure}
\centering
\includegraphics[width = .7\textwidth]{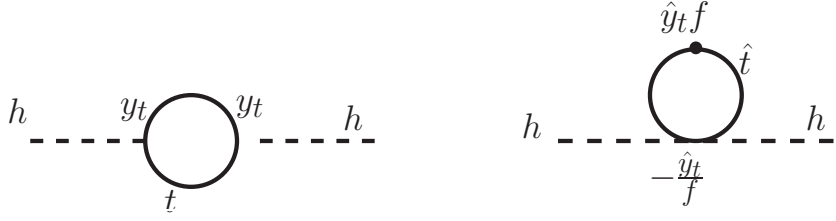}
\caption{Cancellation of the top divergence in the Twin Higgs model. The effective vertex in the second diagram arises upon integrating out the heavy radial mode.}
\label{fig:topcanc}
\end{figure}

We begin with the top yukawa itself. We have seen schematically that the Higgs is protected against large cutoff sensitivity from top loops provided a twin top with $\hyt = y_t$.  But how much can the top Yukawa $\mathbb{Z}_2$ be relaxed while preserving the naturalness of the weak scale?  
When the 
coupling~(\ref{eq:twintopcouple}) is introduced, by Eq.~(\ref{eq:oneloop}) the physical mass of the pseudo-Goldstone boson Higgs gets a quadratically divergent radiative correction at the scale $\Lambda $,
\beq\label{eq:topcancellation}
\delta m_h^2 \approx \frac{3\Lambda^2}{4 \pi^2} \left(  y_t^2 -  \hyt^2\right) \,. 
\eeq 
This precisely cancels out when the ${\mathbb Z}_2$ symmetry is exact.   
We can picture this in Fig.~\ref{fig:topcanc}, where the pseudo-Goldstone Higgs acquires an effective coupling to the twin top upon integrating out the heavy radial Higgs mode. The cancellation in Fig.~\ref{fig:topcanc} is very similar to that in Little Higgs 
theories~\cite{ArkaniHamed:2001nc,ArkaniHamed:2002pa,ArkaniHamed:2002qx},\footnote{See also~\cite{Schmaltz:2005ky} 
for review and references therein.} 
with the difference that the top partner is uncolored.
But without exact ${\mathbb Z}_2$ symmetry, the naturalness demand that these corrections are not much larger than the observed SM-like Higgs mass-squared of $(125~ {\rm GeV})^2$, translates 
into 
\beq \label{dytrequirement}
\left| \frac{\hyt(\Lambda) - y_t(\Lambda)}{y_t(\Lambda)} \right| \lesssim 0.01 \,, 
\eeq 
for $\Lambda \sim 5$ TeV. 

Ordered by the size of tree-level couplings to the Higgs doublet, the next ingredients to consider are the potential twin sector equivalents of $SU(2)_L \times U(1)_Y$ gauge bosons. The contribution to the Higgs mass-squared  from $SU(2)_L$ boson loops in the SM is $\sim (400$~GeV)$^2$ for a cutoff $\Lambda \sim 5$ TeV --- still a significant source of electroweak destabilization, although subdominant to top loops. This suggests gauging the twin $SU(2)$ global symmetry acting on $B$ and $\hQ$. Introducing twin weak gauge bosons with coupling $\hg_2$ translates to the quadratic cutoff sensitivity in the pseudo-Goldstone Higgs,
\beq\label{eq:wcancellation}
\delta m_h^2 \approx \frac{9 \Lambda^2}{32 \pi^2} \left( \hg_2^2(\Lambda) -  g_2^2(\Lambda) \right) \,, 
\eeq 
in analogy with Eq.~(\ref{eq:topcancellation}).  
Demanding this not significantly exceed the observed Higgs mass-squared implies  
 $\left| \frac{\hg_2(\Lambda) - g_2(\Lambda)}{g_2(\Lambda)} \right| \lesssim 0.1$.  Note that with this gauging of twin $SU(2)$ all Goldstone bosons of $SU(4)$ breaking, except for the SM-like Higgs itself, are now longitudinal weak bosons of the visible and twin sectors.

In contrast, the contribution to $m_h^2$ from $U(1)_Y$ loops in the SM is comparable to $m_h^2$ for a cutoff $\Lambda \sim 5$ TeV, and thus already consistent with naturalness.  Thus naturalness does not require twin hypercharge, although it was included in the original Twin Higgs~\cite{Chacko:2005pe}.  This is analogous to the statement that in natural supersymmetry there is no need for the Bino to be light; its presence in the low-energy spectrum is non-minimal from the bottom-up point of view. Given that our principle in this paper is to seek the most economical version of the twin Higgs that is consistent with the naturalness of the little hierarchy, we do not include twin hypercharge in the minimal twin Higgs model, assuming instead that it was never gauged or that it was broken at or around the scale $\Lambda$. However, for the sake of completeness, we will briefly discuss the significant phenomenological consequences of a light twin hypercharge boson in Appendix~\ref{app:hypercharge}.

Next we turn to the twin analogue of QCD. Of course the Higgs does not couple to $SU(3)$ at tree level, but rather at one loop via its coupling to the top quark. This nonetheless leads to sizable two-loop corrections to the Higgs mass from physics around the cutoff. As we will discuss in detail 
in Section~\ref{subsec:perttwinQCD}, the contribution to the Higgs mass-squared from two-loop $y_t^2 g_3^2$ corrections in the SM is {\it at least} $\sim (350$ GeV)$^2$ for a cutoff $\Lambda \sim 5$ TeV, putting QCD on similar footing as the weak gauge group. Gauging the twin $SU(3)$ global symmetry with coupling $\hat{g}_3$ gives quadratic cutoff sensitivity in the pseudo-Goldstone Higgs 
\beq
\delta m_h^2 \approx \frac{3 y_t^2 \Lambda^2}{4 \pi^4} (g_3^2 - \hat{g}_3^2) \,.
\eeq
This is a key observation that will drive the phenomenology of a viable Twin Higgs model: naturalness and minimality favor a confining gauge symmetry in the hidden sector, ``twin glue''.
This twin glue has a coupling close to the QCD coupling  --- we will see how close in Section~\ref{subsec:perttwinQCD} --- and therefore it confines at a scale $\hL$ which is logarithmically close to the SM QCD confinement scale. 

Once the twin $SU(2)$ and $SU(3)$ are gauged, we should include a variety of twin fermions in addition to the twin top quark to cancel anomalies. To render the twin $SU(3)$ anomaly-free we should include a twin RH bottom quark $\hb$; symmetries then admit the hidden sector bottom Yukawa coupling
\beq
\cL = \hat y_b B \hQ \hb \,. 
\eeq
Unlike the case of the top sector,  we should not 
necessarily demand that $\hat y_b = y_b$ since the bottom Yukawa has a much weaker effect on the SM Higgs mass at one loop.   At this stage it suffices that $\hat y_b \ll y_t$ in order to avoid creating a hierarchy problem from the bottom sector. 

Similarly, canceling the twin $SU(2)$ anomaly requires an additional doublet neutral under the twin $SU(3)$: $\hat L$, left-handed twin tau. Although not required for anomaly cancellation, introducing a twin right-handed tau $\hat \tau_R$ allows the $\hat\tau$ to be rendered massive provided a Yukawa coupling
\beq
\cL \supset \hat y_{\tau} B \hat L \hat \tau_R \,.  
\eeq 
As in the case of the twin bottom Yukawa, $\hat y_\tau$ need not respect the ${\mathbb Z}_2$ as long as $\hat y_\tau \ll \hyt$. The twin neutrino may be rendered massive in the same way as the SM neutrinos;  its mass plays no role in naturalness and is essentially a free parameter of the model. Finally, twin light-flavor (first and second generation)  fermions  are totally unnecessary for naturalness, as their Yukawa couplings too small to meaningfully disturb the Higgs potential,  and are therefore absent in our minimal ``Fraternal'' model. 

\subsection{Summary of Fraternal Twin Higgs model}
\label{subsec:summaryofmodel}

Thus we arrive at the ingredients of our minimal Twin Higgs model:
\begin{enumerate}
\item An additional twin Higgs doublet and an approximately $SU(4)$-symmetric potential.
\item Twin tops and a twin top Yukawa that is numerically very close to the SM top Yukawa.
\item Twin weak bosons from the gauged $SU(2)$ with $\hat g_2(\Lambda) \approx g_2(\Lambda)$.
\item Twin glue, a gauged $SU(3)$ symmetry with $\hat g_3(\Lambda) \approx g_3(\Lambda)$. This gauge group
is asymptotically free and it confines, at a scale $\hL$  that we will discuss in Section~\ref{subsec:fratconfinemen}.. 
\item Twin bottoms and twin taus, whose masses are essentially free parameters so long as they remain  much  lighter than the twin top.
\item Twin neutrino from the twin tau doublet, which may have a Majorana mass,  again a free parameter as long as it is sufficiently light.
\end{enumerate}

As an aside, we note that in contrast to a model with a perfect twin of the Standard Model, this model is cosmologically safe; with at worst one massless particle (the twin tau neutrino), and fewer degrees of freedom than the visible sector, the effective number of degrees of freedom during nucleosynthesis and recombination is very small.

Not accidentally, the most crucial ingredients for the naturalness of the theory --- a twin Higgs, twin tops, and twin glue --- strongly resemble the ingredients of natural supersymmetry (Higgsinos, stops, and the gluino).  But the key difference here is the likely existence of a new,  confining gauge group in the minimal twin sector. Although twin glue does not impact the Higgs directly at one loop, its contributions at two loops make it a key component of a viable twin Higgs model.


\section{Electroweak Breaking and Tuning}
\label{sec:tuning}

In this section, we study the effective potential of the Fraternal Twin Higgs model outlined above. We show how realistic  electroweak symmetry breaking can be achieved, accompanied by a $125$ GeV Higgs scalar, 
and estimate the tuning of couplings needed.  In subsection \ref{effpot}, we write down the effective potential for the full Higgs sector of the model at one-loop order.  In subsection \ref{match}, we integrate out the heavier Higgs to get an effective potential for just the SM-like Higgs. In subsection \ref{Tune}, we determine that the degree of fine-tuning needed to have realistic electroweak breaking and SM-like Higgs mass is $\sim 2v^2/f^2$.
Finally, in subsection \ref{justify} we more
fully justify the form of our starting effective potential in subsection \ref{effpot}, by showing that it is free from other types of fine-tuning.

\subsection{Effective Potential}
\label{effpot}

The Twin Higgs effective potential is given, to good approximation, by 
\begin{eqnarray}
V_{eff} &=& \lambda (|A|^2 + |B|^2 - f^2/2)^2 \label{su4} \\
&+&  \kappa (|A|^4 + |B|^4) \label{z2} \\
&+& \rho |A|^4 + \sigma f^2 |A|^2 \label{z2break} \\
&+& \frac{3 y_t^4}{16 \pi^2} \left[ |A|^4 \ln(\Lambda^2/y_t^2 |A|^2) + |B|^4 \ln(\Lambda^2/y_t^2 |B|^2) \right] \label{tloop} \, ,
\end{eqnarray}
where for concrete estimates we will take the UV cutoff of the Twin Higgs theory to be 
$\Lambda = 5 \,{\rm TeV}$.
Lines (\ref{su4}) -- (\ref{z2break}) represent the most general renormalizable tree-level terms consistent with the SM and twin gauge symmetries.  Line (\ref{su4}) is just our starting point,  Eq.~(\ref{eq:simple}), the subset of terms respecting the global $SU(4)$ symmetry, 
 under which $(A, B)$ transform in the  fundamental representation.  Line (\ref{z2}) consists of the extra terms allowed by breaking $SU(4)$ but preserving the discrete ${\mathbb Z}_2$ global subgroup, $A \leftrightarrow B$. (The other ${\mathbb Z}_2$  invariant, $|A|^2|B|^2$, is equivalent to this, modulo $SU(4)$-invariant terms.) 
 Line (\ref{z2break}) consists of the remaining extra terms which respect only the gauge symmetries. Line (\ref{tloop}) is the dominant one-loop radiative correction
  that cannot be absorbed into a redefinition of 
   the tree potential, due to a (twin) top loop. This is just Eq.~(\ref{eq:oneloop}), where we have set $\hat{y}_t = y_t$; in Section~\ref{sec:color}, we will justify this as a good approximation because of the gauging of twin color. 
While the logarithmic cutoff dependence above can be removed by renormalization of $\kappa$, we will keep the above form for later convenience in making estimates.  

\subsection{Matching to SM Effective Field Theory}
\label{match}

Before fully justifying the above approximate structure for the effective potential, we will first work out its  consequences, in particular matching it to a SM effective field theory at lower energies. This will allow us to choose the rough sizes of the different couplings needed for realism, and to then self-consistently check our approximation. 

Line~(\ref{su4}) contains the dominant mass scale, set by $f$. 
We take the $SU(4)$-symmetric self-interaction to be modestly weak, $\lambda \lesssim 1$. 
We seek to stabilize the vacuum such that
\begin{equation}
f =  {\rm few} \times v_A,  ~  ~ v_A =  v  \equiv 246~ {\rm GeV} ,
\end{equation}
so that the pseudo-Goldstone Higgs is primarily SM-like.
We will also  use this small hierarchy to work perturbatively in  powers of $v/f$.

We begin by studying the limit $v/f =0$. 
The dominant potential, (\ref{su4}), then gives rise to a spontaneous symmetry breaking $SU(4) \rightarrow 
SU(3)$, where  the breaking VEV is in the $B$ direction,
\begin{equation}
\langle B_0 \rangle = f/\sqrt{2} \,.
\end{equation}
This breaks twin $SU(2)$ gauge symmetry, with three of the seven Nambu-Goldstone bosons being eaten in the process. The remaining four real Nambu-Goldstone bosons form the complex $A$ doublet, namely the SM Higgs doublet of electroweak gauge symmetry. 
The fourth uneaten $B$ scalar is the radial mode of the potential, a physical heavy exotic Higgs particle with mass,
\begin{equation}
m_{\hat{h}} =  \sqrt{2 \lambda} f \, .
\end{equation} 
Later,  ${\cal O}(v^2/f^2)$ perturbations will  mix these boson identifications to a small extent.

The couplings $\rho$ and $\sigma$ break the discrete ${\mathbb Z}_2$ symmetry. 
The only other couplings in the theory that break ${\mathbb Z}_2$ are the small SM hypercharge coupling and the very small Yukawa couplings to the light fermions (and their twin equivalents in the case of the $\tau$ and $b$, also taken to be $\ll 1$). Therefore it is technically natural to take $\rho$ to be as small as $\sim g_1^4/(16 \pi^2)$, where $g_1$ is the SM hypercharge coupling. This is so small that we neglect it in what follows.\footnote{The possibility of larger $\rho$ offers alternate model building opportunities but we do not pursue them in this paper.}
We can however  consistently take $\sigma f^2$ to represent an explicit {\it soft} breaking of ${\mathbb Z}_2$ symmetry.  Still, we will take $\sigma < \lambda$ so that line (\ref{su4}) dominates the mass scales as assumed above. 

At energies well below the heavy Higgs mass and heavy twin gauge boson masses, set by $f$,
 $|A|^2 + |B|^2$ is rigidly fixed at the bottom of the potential in line (\ref{su4}),  
\begin{equation} \label{AfromB}
|B|^2 = \frac{f^2}{2} - |A|^2 \, .
\end{equation}
Plugging this into the $(A,B)$ effective potential (and neglecting $\rho$ as discussed above) gives an effective potential for just the lighter $A$ degrees of freedom, 
\begin{eqnarray}
\label{smeff}
V_{eff} (A) &=&   \left[ 2\sigma - 2\kappa -  \frac{3 y_t^4}{8 \pi^2} \left( \ln(\Lambda^2/m_{\hat{t}}^2) - \frac{1}{2} \right)  \right] \frac{m_{\hat{t}}^2}{y_t^2} |A|^2  \nonumber \\
&+& \left[ 2 \kappa + \frac{3 y_t^4}{16 \pi^2} \left( \ln(\Lambda^2/y_t^2 |A|^2) +  \ln(\Lambda^2/m_{\hat{t}}^2) - \frac{3}{2} \right) 
\right] |A|^4 \nonumber \\
&+& {\cal O}(A^6/f^2) \, ,
\end{eqnarray}
where this equation is expressed in terms of the mass of the twin top, or ``top partner",
\begin{equation}
m_{\hat{t}} = y_t \frac{f}{\sqrt{2}} \, ,
\end{equation}
in analogy with the top quark mass. 

Eq.~(\ref{smeff}) has the form of a SM effective potential, with the tree-like $|A|^2, |A|^4$ terms, as well as a top-loop induced 
$|A|^4 \ln(\Lambda^2/y_t^2 |A|^2)$ term.  Successful electroweak symmetry breaking, $\langle A_0 \rangle = 246 \,{\rm GeV}/\sqrt{2}$, and a physical Higgs mass of $125$ GeV, can therefore be arranged by tuning the $|A|^4$ coefficient using $\kappa$, and the $|A|^2$ coefficient using $\sigma$. To estimate the tuning involved, we can neglect the modest $\ln|A|$ modulation of $|A|^4$, and just set the logarithm to its expectation value, 
\begin{eqnarray}
\label{smeff2}
V_{eff} &\approx&  \left[ 2\sigma - 2\kappa -  \frac{3 y_t^4}{4 \pi^2} \left( \ln(\Lambda/m_{\hat{t}}) - \frac{1}{4} \right)  \right]  \frac{m_{\hat{t}}^2}{y_t^2} |A|^2  \nonumber \\
&+& \left[ 2 \kappa + \frac{3 y_t^4}{4 \pi^2} \ln(\Lambda/m_{\hat{t}}) + \frac{3 y_t^4}{8 \pi^2} \left( \ln(m_{\hat{t}}/m_{t}) - \frac{3}{4} \right)
\right] |A|^4 \nonumber \\
&\equiv& -  \lambda_{SM} v^2 |A|^2 + \lambda_{SM} |A|^4 \,,
\end{eqnarray}
where $m_t = y_t \langle A_0 \rangle$. For realistic electroweak scale and physical Higgs mass we require $\lambda_{SM} \approx \frac{1}{8}, 
v = 246$ GeV.

Noting that $\ln(m_{\hat{t}}/m_{t}) - \frac{3}{4} = \ln ({\rm few}) - \frac{3}{4} \sim {\cal O}(1)$, we will neglect this term relative to the $\ln(\Lambda/m_{\hat{t}})$-enhanced top-loop contribution to the Higgs quartic, so that 
\begin{equation}
\label{kappa}
\lambda_{SM} \approx  2 \kappa + \frac{3 y_t^4}{4 \pi^2} \ln(\Lambda/m_{\hat{t}}) \,.
\end{equation}
With $y_t \approx 1$, the top-loop induced quartic coupling is already close to the required value, $\lambda_{SM} \sim 1/8$.  
A higher-order and renormalization-group improved analysis, as in studies of analogous corrections in supersymmetry,  is expected to reduce this radiative correction, the central feature of which can be captured by using a top-Yukawa coupling renormalized at several TeV, where $y_t^4 \approx 1/2$. In any case, the rough sizes of the radiative corrections in the Twin Higgs theory are comparable to the realistic value of 
$\lambda_{SM}$, and so with a $\kappa$ of  comparable magnitude we are able to successfully and naturally fit the observed physical Higgs mass.

As an aside, it is instructive to compare the form of the logarithmically divergent (twin) top-loop contributions  in Eq.~(\ref{smeff}) with the analogous contributions from top/stop loops in weak scale supersymmetry (for large $\tan \beta$ and small stop mixing).  As in supersymmetry, the quadratic divergence in the top-loop contribution to the Higgs potential has been canceled by a top partner mechanism. However, there remains a logarithmically divergent contribution to the Higgs mass-squared. This has precisely the same form as in supersymmetry with the replacement, $m_{stop} \rightarrow m_{\hat{t}}$. We see there are also logarithmically divergent contributions to the Higgs quartic self-coupling from the (twin) top loop.  Again, there are analogous contributions in supersymmetry of the same magnitude, but the stop contribution has the opposite sign due to its Bose statistics. Thus in supersymmetry the logarithmic divergence cancels out here, but there is still a finite logarithmic enhancement factor of $\ln(m_{stop}^2/m_t^2)$ to the Higgs quartic correction. Unlike the MSSM where the tree-level Higgs quartic is set by electroweak gauge couplings, here we have an unconstrained tree-level quartic contribution, $2\kappa$. (However, $\kappa$ also appears in the Higgs mass term, so it will affect  electroweak tuning, as discussed below.)

\subsection{Estimating Electroweak Tuning}
\label{Tune}

Electroweak tuning involves the quadratic terms of the potential.
Fortunately, the same combination of parameters that dominates the quartic self-coupling, in the second line of  Eq.~(\ref{smeff2}), also appears in the quadratic terms of the potential, 
 in the first line of Eq.~(\ref{smeff2}).  
 We will drop the $1/4$ term relative to  
 the $\ln \Lambda$-enhanced part of the top radiative correction. Then, 
 matching the quadratic terms to the SM form on the last line, we have
\begin{equation}
\label{sigma}
\lambda_{SM} v^2 \approx  (\lambda_{SM} - 2 \sigma) \frac{m_{\hat{t}}^2}{y_t^2} \, .
\end{equation}
The degree of electroweak tuning of $\sigma$ needed to achieve this is therefore
\begin{eqnarray}
\label{ewtune}
{\rm Electroweak-Tuning} &\sim& \frac{ \lambda_{SM} v^2}{\lambda_{SM}  (m_{\hat{t}}^2/y_t^2)} \nonumber \\
&=&  \frac{2 m_t^2}{m_{\hat{t}}^2} =  \frac{2 v^2}{f^2} \, .
\end{eqnarray} 
 For example, for  $f \sim 3 v$ ($m_{\hat{t}} \approx 500 - 600$ GeV), this corresponds to a very mild $20$ percent electroweak tuning.

\subsection{Twin Higgs Effective Potential Approximation}
\label{justify}

Finally, we justify our starting approximation for the Twin Higgs effective potential upon which our analysis of electroweak breaking and tuning is based. To do this, we examine each term of Eqs. (\ref{su4} -- \ref{tloop}) and ask whether its coefficient is radiatively stable, and also whether this coupling itself significantly radiatively corrects other couplings. 
 
The tuning needed to have $f$ smaller than the cutoff $\Lambda$ is determined by the leading radiative corrections to the quadratic terms in line (\ref{su4}).  We have taken the $SU(4)$-symmetric self-interaction to be weakly coupled enough that  
  the leading quadratic radiative corrections come from the (twin) top loop and the large Yukawa coupling, 
 \begin{equation}
\label{topquad}
 V^{(1-loop)} \supset \frac{3}{8 \pi^2} y_t^2 \Lambda^2 \left( |A|^2 + |B|^2 \right) \, .
\end{equation}
The other couplings that can similarly contribute at one loop are the electroweak gauge couplings, as well as $\kappa$ and $\rho$. The electroweak couplings are subdominant to the top Yukawa coupling as reviewed earlier. We have estimated that  $\kappa \sim {\cal O}(\lambda_{SM})$   (Eq.~(\ref{kappa}) and ensuing discussion), and loops using the SM Higgs quartic coupling are also considerably subdominant to top loops. Finally, we have consistently chosen $\rho$ to be very small as explained earlier.  Therefore, Eq.~(\ref{topquad}) dominates radiative corrections to $\lambda f^2$ and gives a tuning 
 to keep $f  \ll \Lambda$ of
\begin{eqnarray}
\label{tunef}
f-{\rm tuning} &\sim& \frac{\lambda f^2}{\frac{3}{8 \pi^2} y_t^2 \Lambda^2} =  \frac{8 \pi^2 \lambda f^2}{3 y_t^2 \Lambda^2} \, .
\end{eqnarray} 
If we take for example $f \sim 3 v \sim 750$ GeV, we see that there is essentially no tuning. 

We are taking $\kappa$ to be comparable to the leading radiative corrections from the top loop (Eq.(\ref{kappa}) and ensuing discussion), so  this 
${\mathbb Z}_2$-preserving but $SU(4)$-violating coupling is radiatively stable and natural. As explained earlier, we can naturally take the {\it hard} breaking of ${\mathbb Z}_2$ given by $\rho$ to be negligibly small.  $\sigma f^2$ represents a {\it soft} breaking of ${\mathbb Z}_2$, with $\sigma \sim \lambda_{SM}/2$ (Eq.~(\ref{sigma}) and ensuing discussion), and is clearly natural without other sources of soft breaking. To tune to realistic electroweak symmetry breaking we saw that we needed $\sigma \sim \lambda_{SM}/2 \sim 1/16$, therefore we can easily have $\sigma f^2 \ll \lambda f^2$ to ensure that line (\ref{su4}) indeed dominates the mass scales of the Twin Higgs potential as our analysis presumes.  

With all couplings in our theory being perturbative, our one-loop analysis suffices for demonstrating the successful matching to a realistic SM effective field theory and for estimating the tuning required.


\section{Fraternal Color \label{sec:color}}

 We now discuss the dynamics of twin color in more detail. We first calculate how close the twin and visible sector color couplings must be to preserve naturalness. Then we estimate the confinement scale of twin color and discuss the associated twin hadrons, including glueballs and quarkonia.
 
\subsection{Perturbative Considerations} \label{subsec:perttwinQCD}

As we have seen above, the twin Higgs mechanism for naturalness, at one-loop order, requires the top and twin-top Yukawa couplings to be very nearly identical close to the cutoff. Here we show that when QCD effects are taken into account at two-loop order, naturalness favors having a twin QCD, with a gauge coupling similar to that of QCD near the cutoff $\Lambda$. 

To see this, consider the one-loop RG analysis for the dimensionless Wilsonian running mass-squared parameter of the SM-like Higgs,  $x(\mu) \equiv \frac{m^2_h(\mu)}{\mu^2}$:
\begin{eqnarray}
\label{RGx}
\frac{dx}{d \ln \mu} = - 2 x + \frac{3 (\hat{y}_t^2 - y_t^2)}{2 \pi^2} & ; &\nonumber \\
\frac{d y_t}{d \ln \mu} = \frac{9 y_t^3}{32 \pi^2} - \frac{y_t g_3^2}{2 \pi^2} \ \ &;& \ \
\frac{d \hat{y}_t}{d \ln \mu} = \frac{9 \hat{y}_t^3}{32 \pi^2} - \frac{\hat{y}_t \hat{g}_3^2}{2 \pi^2} \ ; \nonumber \\  
\frac{d g_3}{d \ln \mu} = - \frac{7 g_3^3}{16 \pi^2} \ \ &;& \ \
\frac{d \hat{g}_3}{d \ln \mu} = - \frac{29 \hat{g}_3^3}{48 \pi^2} \, .
\end{eqnarray}
where $g_3$ is the SM QCD coupling and $\hat{g}_3$ is the twin QCD coupling. Note that $g_3$ and $\hat{g}_3$ run differently because 
the SM has six QCD quark flavors while the minimal model has just two twin QCD flavors. 
If we neglect the running of $y_t, \hat{y}_t, g_3, \hat{g}_3$, then the solution to the first of these equations is simply given by 
\begin{equation}
m^2_h(\mu) = m^2_h(\Lambda) + \frac{3 (y_t^2 - \hat{y}_t^2)}{4 \pi^2} (\Lambda^2 - \mu^2) \, .
\end{equation}
In running down to the weak scale or physical Higgs mass, $\mu \ll \Lambda$, we see we have merely matched onto the one-loop result of Eq.~(\ref{eq:topcancellation}), 
which stresses the importance of having $y_t$ and $\hat{y}_t$ be very nearly the same. 
But taking account of the running of  $y_t, \hat{y}_t, g_3, \hat{g}_3$ by solving  all of Eqs. (\ref{RGx}) then gives  an  RG-improved result, which allows us to  explore the role of twin color  in maintaining $y_t\sim\hat{y}_t$ as the couplings run.

The simplest calculation arises by seeing what happens when we have no twin color, setting $\hat{g}_3 = 0$.  The danger to naturalness is then that the SM QCD coupling will give $y_t(\mu)$ a different evolution to $\hat{y}_t(\mu)$ even if they happen to coincide at $\mu = \Lambda$, and that this deviation will feed into $m_h^2$. To focus on just this effect we will drop the $g_3$-independent terms in the $y_t, \hat{y}_t$ $\beta$-functions. since these have identical forms for the twin and SM sectors.  (Keeping these effects would be subleading in the running of $m_h^2$  compared to those of the $g_3$-dependent term if we start with $y_t(\Lambda) \approx \hat{y}_t(\Lambda)$.)
 As a final simplifying approximation we will drop the running of $g_3$ itself, because its $\beta$-function is small near $\Lambda$ and because we will see that the running in $m_h^2$ is dominated near $\Lambda$. With these approximations, and working to first order in $\hat y_t - y_t$, the solution to Eqs. (\ref{RGx}) is given by
\begin{equation} 
m_h^2(\mu) \approx m_h^2(\Lambda) + \left[\frac{3 y_t^2 g_3^2}{8 \pi^4} +  \frac{3 (y_t^2 - \hat{y}_t^2)}{4 \pi^2} \right](\Lambda^2 - \mu^2) 
+  \frac{3 y_t^2 g_3^2}{4 \pi^4} \mu^2 \ln(\mu/\Lambda)\, .
\end{equation}
Running down to low scales, $\mu \ll  \Lambda$, 
\begin{equation} 
\label{RGxsoln}
m^2_{h,IR} \approx m_h^2(\Lambda) + \left[\frac{3 y_t^2 g_3^2}{8 \pi^4} +  \frac{3 (y^2_t - \hat{y}^2_t)}{4 \pi^2} \right] \Lambda^2 \, .
\end{equation}
In the above equations  all the dimensionless couplings are evaluated at $\Lambda$.

We see that even if $y_t(\Lambda) = \hat{y}_t(\Lambda)$,   just the running from SM QCD has led to a quadratic divergence which would require a fine tuning of the counter-term $m_h^2(\Lambda)$ to get the physical Higgs mass,  
\begin{equation}
{\rm Two~loop ~fine~tuning} \sim \frac{m^2_{h,{\rm physical}}}{\frac{3 y^2_t(\Lambda) g^2_3(\Lambda)}{8 \pi^4} \Lambda^2} \approx 0.25.
\end{equation}
Of course, it is not reasonable for two couplings, $y_t, \hat{y}_t$, that run differently to be exactly the same at $\Lambda$. This would be a fine tuning in itself.  A technically natural estimate is that they differ by at least the running of $y_t$ due to QCD over an e-folding of running, 
\begin{equation} \label{threshold}
|y_t(\Lambda) - \hat{y}_t(\Lambda)| \gtrsim \frac{g^2_3(\Lambda)}{2 \pi^2} y_t(\Lambda) \, .
\end{equation} 
In other words, we expect a comparable quadratic divergence in $m^2_{h,IR}$ from the splitting in Yukawa couplings and from  the explicit ${\cal O}(g_3^2 y_t^2)$ divergence. 
Thus a better estimate of fine-tuning {\it in the absence of twin QCD} is $\lesssim 10$ percent. 

The tuning due to QCD at two-loop order will clearly become very mild if we do include twin QCD with $\hat{g}_3(\Lambda) \approx g_3(\Lambda)$. The two couplings will run differently in the IR due to the different number of SM and twin generations, but quadratic sensitivity to $\Lambda$ 
will be determined by the UV couplings. Roughly, Eq.~(\ref{RGxsoln}) is then replaced by 
\begin{equation} 
\label{RGxsoln2}
m^2_{h, IR} \approx m_h^2(\Lambda) + \left[\frac{3 y_t^2(\Lambda) (g_3^2(\Lambda) - \hat{g}_3^2(\Lambda))}{8 \pi^4} +  \frac{3 (y^2_t(\Lambda) - \hat{y}^2_t(\Lambda))}{4 \pi^2} \right]\Lambda^2 \, .
\end{equation}
Therefore, if $g_3(\Lambda)$ and $\hat{g}_3(\Lambda)$ agree to within even $15 \%$, enforced by an approximate ${\mathbb Z}_2$, the tuning in $m_h^2$ will be a mild $\sim 30 \%$, comparable to that in Eq.~(\ref{ewtune}). This estimate combines the $\mathcal{O}(g_3^2 y_t^2)$ contribution from running and the threshold correction in Eq.~(\ref{threshold}) in quadrature to reflect the unknown relative sign of the threshold correction. We will take this to be the case in what follows and study the effects of twin confinement from this twin QCD sector on the twin spectrum and phenomenology. 

Finally, of course even $g_3$ and $\hat{g}_3$ run differently because the particle content of the {\it minimal} Twin sector differs from that of the SM, so one may wonder how close they can naturally be at $\Lambda$. However, the analogous estimate to Eq. (\ref{threshold}) is 
\begin{equation} 
\frac{|g_3(\Lambda) - \hat{g}_3(\Lambda)|}{g_3(\Lambda)} \gtrsim \frac{g^2_3(\Lambda)}{6 \pi^2}  \, ,
\end{equation} 
which is easily consistent with the requirement of naturalness discussed above.

\subsection{Fraternal Confinement}
\label{subsec:fratconfinemen}

We now determine the confinement scale $\hat \Lambda_3$ of the twin $SU(3)$ gauge interaction, as this governs the infrared phenomenology of the twin sector.

If all fermions carrying twin $SU(3)$ quantum numbers are much heavier than the confinement scale, the infrared physics is that of pure $SU(3)$ gauge fields, and the lightest states in 
the confined twin sector will be glueballs, whose rich spectrum includes states with different angular momentum $J$, charge conjugation 
$C$ and parity $P$.  
As shown in lattice studies of $SU(3)$ pure glue~\cite{Morningstar:1999rf,Chen:2005mg}, 
at least a dozen glueballs are stable against decay to other glueballs, and the lightest, which we will refer to as $\zpp$, has $J^{PC} = 0^{++}$.   Lattice data provides the ratios of these glueballs' masses to each other and to the confinement scale.  To determine the physics of the twin sector thus only requires us
to compute $\hat \Lambda_3$. 

Once the twin $b$ becomes sufficiently light that twin glueballs and twin bottomonium states (which we will refer to generically as ``$\bbtwin$'') have comparable masses, the situation becomes more complex.    We will not explore this regime carefully in this paper, leaving its details for future study.  However, the calculation of $\hat\Lambda_3$ and of glueball masses given below still applies approximately.

\begin{figure}[t] 
   \centering
          \includegraphics[width=3in]{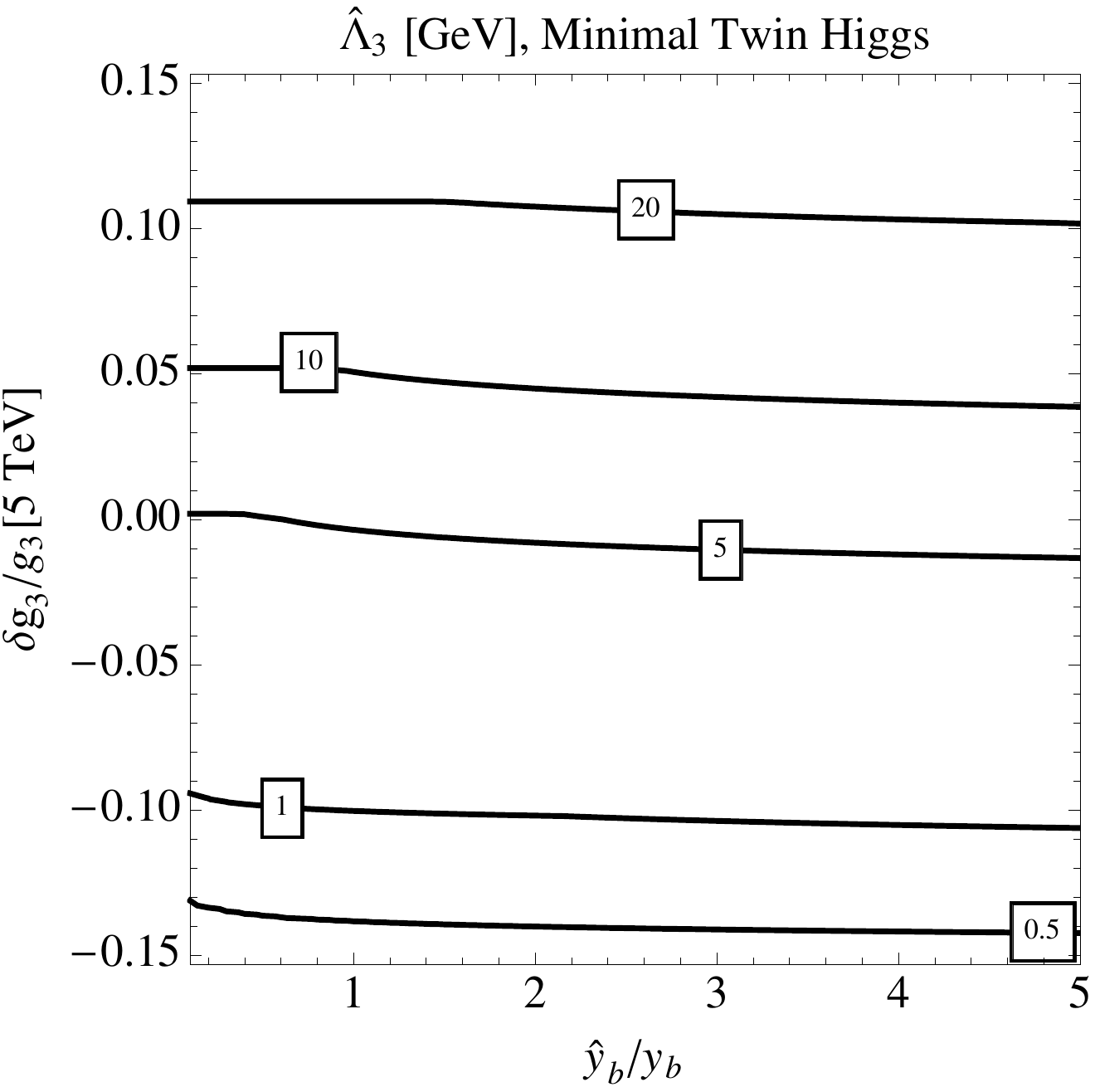}
   \caption{The confinement scale $\hat \Lambda_3$ of the twin $SU(3)$ coupling given fractional variations in $\hat g_3$ and $\hat y_b$ at the cutoff $\Lambda$ for the minimal Twin Higgs (dependence on $\hat y_t$ is negligible). Here we take $\Lambda = 10 \hat m_t$ and $f = 3v$. The mild kinks are due to the $\hat b$ threshold.}
   \label{fig:confinement}
\end{figure}

The twin and SM $SU(3)$ couplings are similar at the cutoff $\Lambda$, but the twin sector has fewer quark flavors, faster running (i.e.~a more negative beta function), and therefore a modestly higher confinement scale.   This is illustrated in Fig.~\ref{fig:confinement}, which shows the strong coupling scale $\hL$ of twin QCD as a function of the variation $\delta g_3$ between SM and twin QCD couplings at the cutoff as well as the value of $\hat y_b$ relative to $y_b$. Note that for $g_3 \approx \hat g_3$, $\hL$ is typically one to two orders of magnitude above that of QCD, with weak dependence on $\hat y_b$ through its impact on the twin QCD beta function.\footnote{At two loops, we define the $\overline{MS}$ confinement scale $\hat \Lambda_3$ via
\begin{equation}
\frac{\hat \Lambda_3^{\overline{MS}}}{\mu} = \exp \left( - \frac{1}{2 b_0 \hat g_3^2(\mu)} \right) \left( b_0 \hat g_3^2(\mu) \right)^{-b_1/2 b_0^2} \left( 1 + \frac{b_1}{b_0} \hat g_3^2(\mu) \right)^{b_1/2 b_0^2} \, ,
\end{equation}
where $b_0, b_1$ are the one-loop and two-loop twin QCD beta functions respectively,
and $\hat g_3(\mu)$ is understood to be the $\overline{MS}$ coupling. }

\begin{figure}[t] 
   \centering
          \includegraphics[width=5in]{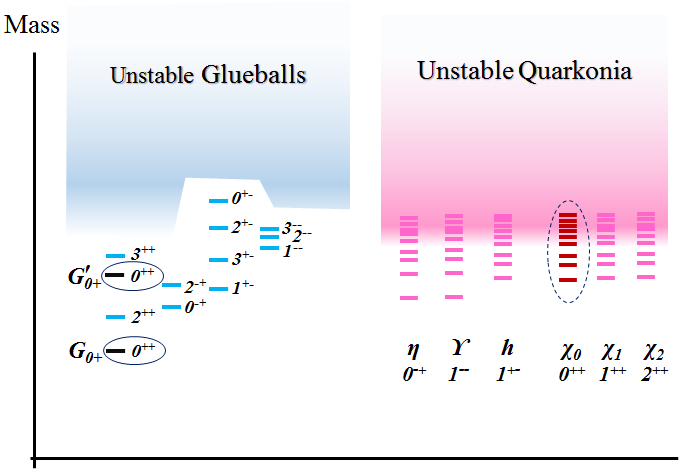}
          \caption{Sketch of the twin hadron spectrum in the regime where $m_0<2m_\btwin<2m_0$.  In addition to the $\zpp$, of mass $m_0$, about a dozen other glueballs, with mass splittings of order $m_0$, are stable against twin strong decays.  Numerous twin bottomonium states, including a tower of $0^{++}$ states $\chib$, are stable against twin strong decays.   The circled $\zpp$ and $G'_{0+}$ glueballs, and potentially the $\chib$ quarkonia, can dominantly decay via annihilation through an $s$-channel  off-shell Higgs to the SM.}
   \label{fig:HadronSpectrum}
\end{figure}

We may now estimate the mass scale of twin glueballs. Using lattice estimates of the glueball mass spectrum in units of the inverse force radius~\cite{Morningstar:1999rf,Chen:2005mg} and the zero-flavor $SU(3)$ $\overline{MS}$ confinement scale in units of the inverse force 
radius~\cite{Gockeler:2005rv}, we find the mass $m_0$ of the $\zpp$ glueball to be related to the strong coupling scale via $m_0 = 6.8 \hL^{\overline{MS}}$. The physical scale $m_{0}$ determined by running couplings down from the cutoff carries a combined uncertainty of $\mathcal{O}(10\%)$ from the lattice estimates, primarily due to uncertainty in the inverse force radius~\cite{Gockeler:2005rv}. Given the mass of the $\zpp$, the masses of higher glueball excitations in terms of $m_0$ are known to good precision. The next highest states in the glueball spectrum are well separated from the $\zpp$, with the closest states being the $G_{2+}$, a $2^{++}$ glueball with $m_{2++} \sim 1.4\ m_0$, and the $\zmp$, a $0^{-+}$ state with $m_{0-+} \sim 1.5\ m_0$~\cite{Morningstar:1999rf,Chen:2005mg}.  It appears also that there is a second stable $0^{++}$ glueball, the $G'_{0+}$, with 
$m\sim 1.8\ m_0$~\cite{Lucini:2004my}. The twin glueball spectrum is illustrated on the left-hand side of Fig.~\ref{fig:HadronSpectrum}.

Meanwhile, the twin bottomonium states form a rich spectrum, whose lowest lying states are narrow if rapid decays via twin glueballs are inaccessible.  As is familiar from SM quarkonium data, and as sketched on the right-hand side of Fig.~\ref{fig:HadronSpectrum}, the spectrum  includes towers of $0^{+-}$, $1^{--}$ and $j^{++}$ ($j=0,1,2$) states, which by analogy we call $\etab$, $\hat \Upsilon$ and $\chib_j$.  The lowest-lying twin quarkonia states have masses of order $2m_\btwin$, with mass splittings $\ll 2m_\btwin$ (as long as $m_\btwin\gg\hat\Lambda_3$.)  However, since there are no other light twin quarks, there are no ``open twin bottom'' mesons analogous to the SM's $B_u, B_d$ mesons.  Thus the towers of narrow quarkonium states extend much further up than in the SM, potentially up to a scale of order $2m_0$, as sketched in Fig.~\ref{fig:HadronSpectrum}, or $m_0+2m_\btwin$.

The reader should note that Fig.~\ref{fig:HadronSpectrum} is only illustrative, and must be interpreted with caution.  Its details change dramatically as one raises or lowers the quarkonium masses relative to the glueball masses, and it omits the many narrow higher-spin quarkonium states, along with various other phenomenological details.

\section{Twin Hadron Phenomenology}
\label{sec:twinhadronpheno}

Thus far we have seen that viable Fraternal Twin Higgs models include twin glue, with couplings that favor confinement 
roughly an order of magnitude or so larger than the SM QCD scale, producing relatively light twin glueballs and/or twin quarkonia.  Both twin gluons and twin quarks are connected to the Standard Model via low-dimensional portals, and this can lead to  observable and even spectacular twin hadron phenomenology. As in Folded Supersymmetry~\cite{Burdman:2006tz}, where twin glueballs also arise, we thus find a connection between dark naturalness and twin hadrons.  In our case, this connection manifests itself as new and exciting opportunities for discovery at the LHC. 

The model's phenomenology changes significantly as we move around in
the parameter space, and in most regions it is rather complicated.
But the most promising and dramatic LHC signals arise even in the
conceptually simplest region, namely where $m_\btwin>\frac{1}{2}m_h$ (i.e., in
Fig.~\ref{fig:Regions0} below, the part of region A above the dashed line).  In this case
the main phenomenon is that described in Fig.~\ref{fig:makingglueballs}, with twin gluons
produced in $h$ decays and hadronizing into twin glueballs, including
the $\zpp$.   The $\zpp$ lifetime is discussed in Section~\ref{subsec:ggdecay},
Eqs.~(\ref{eq:zpp2YY}) - (\ref{eq:GlueballLifetime}); the (perturbative) production rate for glueballs is
discussed in
Section~\ref{subsec:twinhadronproduc}, Eq.~(\ref{eq:IrreducibleWidth}), with nonperturbative subtleties described
in Appendix~\ref{subapp:nonperthgg}.
The reader seeking to avoid becoming lost
in details at a first reading may wish to focus merely on this simple
scenario, in which case Section~\ref{subsec:bbdecay}, the later portions of Section \ref{subsec:twinhadronproduc}
, and Appendices~\ref{app:quarkonium} and \ref{subapp:nonperthbb} may be omitted.

\begin{figure}[t] 
   \centering
          \includegraphics[width=4in]{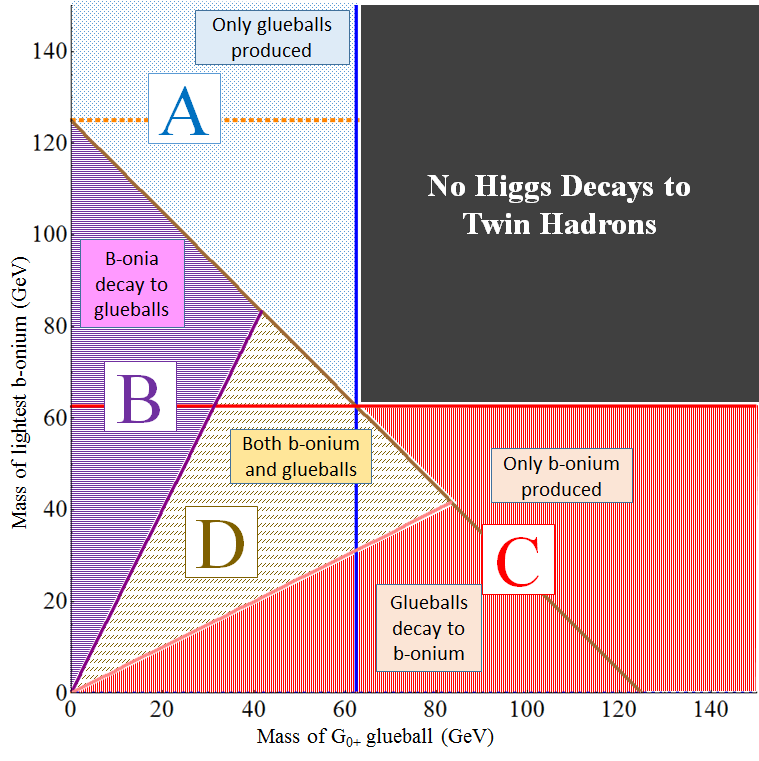}
          \caption{The parameter space of the model in terms of the masses of the lightest  glueball $\zpp$ and the lightest quarkonium $\etab$.  In region A, only glueballs are produced; in region B, the relevant quarkonia decay to glueballs;  in region C, glueballs are either not produced or decay to quarkonia, so only quarkonia appear in the final state; and in region D there are both metastable glueballs and metastable quarkonia, with the potential for mixing.  Solid lines indicate kinematic boundaries.
          }
   \label{fig:Regions0}
\end{figure}

\subsection{Kinematic Regions}

  Before we begin, it is useful to parameterize the theory through $m_0$ and $m_\etab$ (as well as $f$) in place of $\hat g_3,\hat y_b$.  Here $\etab$ is the lightest $\bbtwin$ state, lying slightly below the lightest $\chib$ state.  We can then divide the parameter space of the model into four qualitatively different kinematic regions, shown in Fig.~\ref{fig:Regions0}:
  \begin{itemize}\item Region A: $m_h>2m_0$, $m_h<2m_\etab$ and $m_h<m_0+m_\etab$, so that $h$ can decay to twin glueballs but not to twin bottomonium.
  \item Region B: $m_h>m_0+m_\etab$ and $m_\etab>2m_0$; here $h$ can produce twin bottomonium, but all $\chib$ states (and most or all other $\bbtwin$ states) decay eventually to twin glueballs.
  \item Region C: Either $m_0+m_\etab>m_h>2m_\etab$, in which case only twin bottomonium can be produced in $h$ decays, or $m_0>2m_\etab$, so that any produced glueballs decay rapidly to twin bottomonium.\footnote{Not all of region C is physically meaningful.  As $m_0$ rises, so does $\hL$, and $m_\etab$ always remains heavier than the confinement scale.} Only bottomonia appear 
  in the final states. 
  \item Region D: $m_h>m_0+m_\etab$, $m_\etab<2m_0$, $m_0>2m_\etab$; here both the lighter twin bottomonium and glueball states are metastable, and either or both may appear in final states.  Mixing between the two classes of states can be important in this region.
  \end{itemize}
As we will see, visible decays are typical, and displaced decays are possible, in regions A, B and D.  They may or may not be absent (or very rare) in most or all of region C; this is model-dependent.

\subsection{Couplings to the Visible Sector}
\label{subsec:portals}

In this model, the two portals of greatest importance will involve the dimension-five operator $A^\dag A \hat b \hat{\bar{b}}$ and  the dimension-six operator $A^\dag A \, \hat G_{\mu \nu} \hat G^{\mu \nu}$, where recall $A$ is the SM-like Higgs doublet.
After SM electroweak symmetry breaking and twin confinement, the first operator causes mixing between the SM-like Higgs and twin $\hat \chi$ quarkonia, while the second operator causes mixing between the SM-like Higgs and twin $\zpp$ glueballs.
The effective dimension-5 coupling $A^\dag A \hat b \hat{\bar{b}}$ originates from the twin bottom Yukawa $\hat{y}_b B \hat Q \hat d$; applying (\ref{AfromB}) gives the leading interaction 
\begin{equation}
\mathcal{L}_5 = - \hat{y}_b \frac{A^\dag A}{\sqrt{2} f} \hat b \hat{\bar{b}} \, ,
\end{equation}
which after electroweak symmetry breaking yields $\mathcal{L} \supset - \frac{\hat{y}_b}{\sqrt{2}} \frac{v}{ f} h \hat b \hat{\bar{b}}$. This is just the $v/f$-suppressed coupling of the SM-like Higgs to twin fermions, which gives rise to mixing between $h$ and $\hat \chi$ quarkonia.\footnote{
No other portals are relevant for $h$ decays.  Below the heavy
$\hat h$ scale, and in the absence of a twin $U(1)$, there are no
dimension-4 portals; however we do discuss the effect of these portals
in Section~\ref{subsec:heavyhiggs} and Appendix~\ref{app:hypercharge}. A dimension-5 portal involving SM and twin neutrino mixing is not induced or required in our model and may
easily be too small to have any measurable effect.} 

 While the value of the $h - \hat \chi$ mixing can vary depending on the (unknown) value of the twin bottom Yukawa, the $h-\zpp$ coupling is necessarily generated by the basic ingredients of the minimal Twin Higgs.  In a manner  entirely analogous to the Standard Model $hgg$ coupling, loops of twin tops generate an effective coupling between the twin Higgs doublet $B$ and twin gluons; after $SU(4)$ breaking and SM electroweak symmetry breaking, this leads to a coupling between twin glue and the SM-like Higgs. The effective coupling between the twin Higgs doublet $B$ and twin glue takes the usual form
 \begin{equation}
\mathcal{L}_6 = \frac{\hat \alpha_{3}}{12 \pi} \hat G^a_{\mu \nu} \hat G_a^{\mu \nu} \ln \left( \frac{B^\dag B }{f^2} \right) \, ,
\end{equation}
and applying (\ref{AfromB}) generates the corresponding coupling to $A^\dag A$, which after electroweak symmetry breaking leads to
\begin{equation} \label{eq:dim6}
\mathcal{L} \supset- \frac{\hat \alpha_{3}}{6 \pi} \frac{v}{f} \frac{h}{f} \hat G^a_{\mu \nu} \hat G_a^{\mu \nu}  \, .
\end{equation}
Here the couplings are largely fixed by naturalness considerations.

\subsection{Glueball Decay}
\label{subsec:ggdecay}

Once produced, twin glueballs can decay into kinematically available final states. Assuming there are no light quarks in the twin sector, the only potentially available decays are into light Standard Model fermions via the (off-shell) SM-like Higgs $h$ or into the twin lepton sector via the heavier Higgs $\hat h$  or twin $\hat Z$. 
The decay
$\zpp \to h^* \to YY$, where $Y$ are light SM fields, provides a visible signal.
This process was studied in~\cite{Juknevich:2009gg} in the context of a similar Hidden Valley model~\cite{Juknevich:2009ji}, with a similar Higgs portal for decays 
of the 
glueballs, and with a production portal induced by ``quirks''~\cite{Strassler:2006im,Kang:2008ea}.
The width for $\zpp  \to YY$ is
\begin{equation}\label{eq:zpp2YY}
\Gamma_{\zpp \to YY} = \left( \frac{\hat \alpha_3 v  {\mathfrak{f}}_0}{6 \pi f^2 (m_h^2 - m_0^2)} \right)^2 \Gamma_{h \to YY}^{SM}(m_0^2) \, .
\end{equation}
Here $ \Gamma_{h \to YY}^{SM}(m_0^2)$ is the width of a SM-like Higgs of mass $m_h = m_0$ and 
$\mathfrak{f}_0$ is the $\zpp$ decay constant; from the lattice we have 
$4 \pi \hat \alpha_3 \mathfrak{f}_0 = 3.06 m_0^3$~\cite{Chen:2005mg}. This is the dominant decay mode 
of twin glueballs in the minimal 
model; decays into twin sector leptons are subleading, due to suppression factors of $(m_h/m_{\hat h})^4(f/v)^2$ for decays via $\hat h$ and by an extra $(v/f)^2$ mixing factor if via $h$.

\begin{figure}[t] 
   \centering
 
     \includegraphics[width=3in]{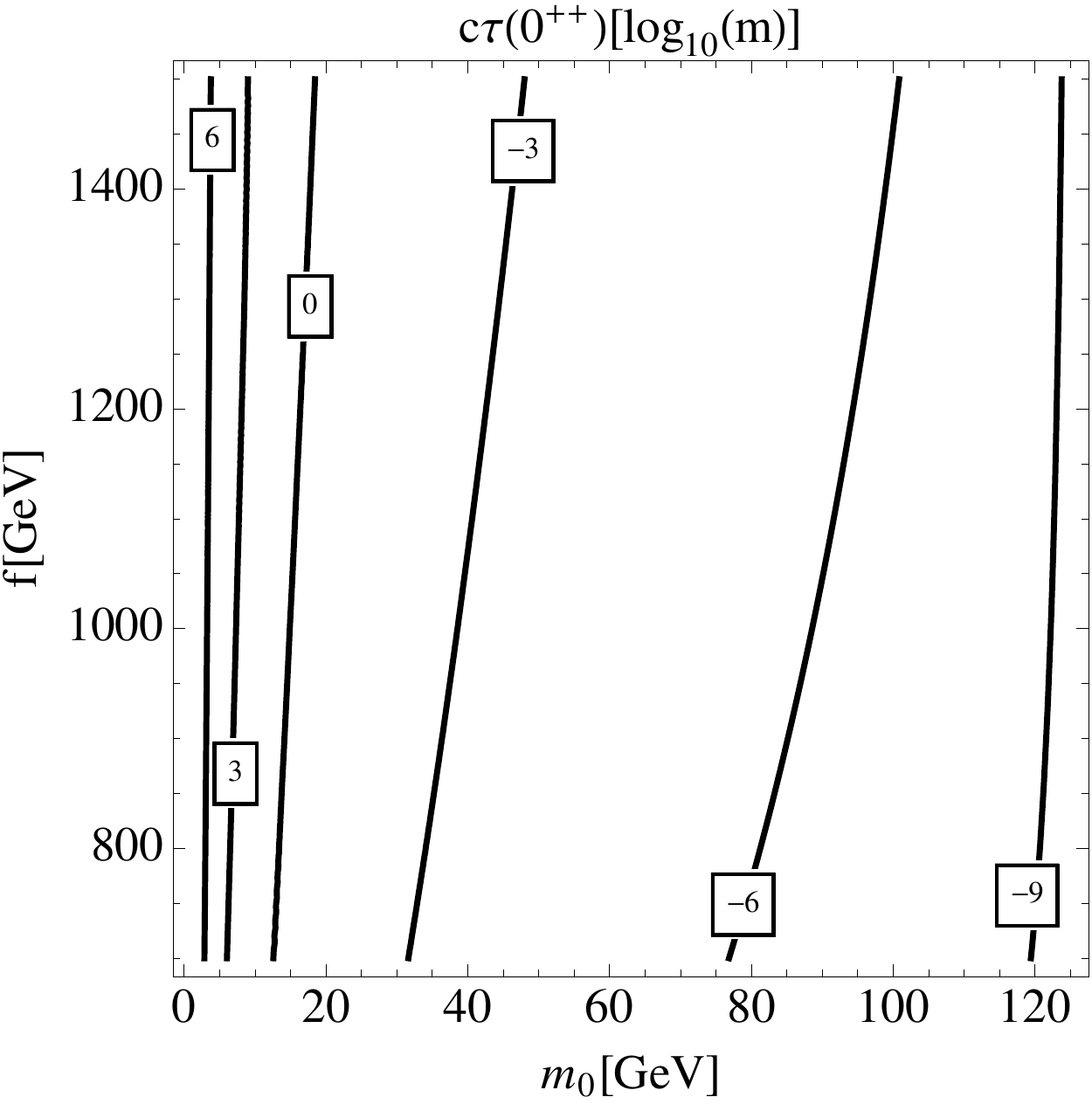} 
   \caption{ Decay length $c\tau_0$ of the $\zpp$ state in $\log_{10}({\rm meters})$ as a function of $m_0$ and $f$. }
   \label{fig:hwidth}
\end{figure}

 Provided that the $\zpp$ decays primarily into SM final states, we can determine its lifetime $\tau_0$ in terms of $m_0$ and $f$. This exercise is particularly straightforward for glueballs much lighter than the massive Standard Model gauge bosons, for which the factor $\Gamma_{h \to YY}^{SM}(m_0^2)$ scales linearly with glueball mass because decays into longitudinally-coupled modes are doubly off-shell. Then in this regime we have simply 
  \begin{eqnarray}\label{eq:GlueballDecayRate}
\Gamma \sim 1.1 \times 10^{-17} \; {\rm GeV} \;\times  \left( \frac{m_0}{10 \; {\rm GeV}} \right)^7 \left(\frac{750 \; {\rm GeV}}{f} \right)^{4}  \, ,
\end{eqnarray}
valid for $2 m_\tau  \lesssim m_0 \lesssim m_W.$ This corresponds to a decay length $c \tau_0$ of approximately
\begin{equation}\label{eq:GlueballLifetime}
c \tau_0 \sim 18 \; {\rm m} \; \times \left( \frac{10 \; {\rm GeV}}{m_0} \right)^7 \left( \frac{f}{750 \; {\rm GeV}} \right)^4 \, .
\end{equation}
Note $m_0 \sim 10-50$ GeV is a central range of values for the glueball mass, given (a) the relative factor of $\sim 7$ relating $\hat \Lambda_3$ to $m_0$ and (b) the higher confinement scale of twin QCD given a reduced number of twin sector fermions, see Fig.~\ref{fig:confinement}.  This is a tantalizing result from an experimental perspective; it implies that the twin sector glueballs give rise to displaced decays on the length scale of the LHC detectors. Even for average decay lengths greater than the scale of the detectors, the large number of $h$ bosons produced at the LHC means that an appreciable number of distinctive displaced decays can still occur within the detector. These decays involve a mix of final states with relative rates corresponding to the decay of an SM-like Higgs of mass $m_h = m_0$ --- primarily bottom quarks, if kinematically accessible, as well as tau pairs and gluons.

We can improve upon the above parametrics by computing the SM Higgs width as a function of mass using \texttt{HDECAY} \cite{Djouadi:1997yw}. A precise plot of the decay length $c\tau_0$ as a function of $m_0$ and $f$ is then shown in Fig.~\ref{fig:hwidth}.  As we can see, the decay length consistent with naturalness in the minimal model is typically in the range of millimeters to meters,
with a jump toward the kilometer scale for $m_0<2m_b$.

Note, however, that given the strong dependence of the glueball decay length on the glueball mass, $c \tau_0 \propto m_0^7$, even modest uncertainties in the value of $m_0$ may substantially influence this estimate. For example, the  $\mathcal{O}(10\%)$ uncertainty on the value of $m_\zpp$ inferred from lattice data translates into a factor of two variation in the value of $c \tau_0$.

The second $0^{++}$ glueball $G'_{0+}$ decays similarly to the first, with a shorter lifetime, and may have prompt or displaced decays. Others such as the $2^{++}$ may decay via an off-shell Higgs, e.g. $G_{2+}\to\zpp h^*$. At small masses their lifetimes are very long \cite{Juknevich:2009gg}, and they will appear only as missing energy in $h$ decays. At larger $m_0$ they cannot appear in $h$ decays, though they might possibly appear as displaced vertices in decays of the heavy twin particles $\hat Z$ or $\hat h$.

The above calculations apply in regions A and B but are irrelevant in region C, where instead $\zpp\to\bbtwin\bbtwin$.  They are only part of the story in region D 
due to glueball-quarkonium mixing, which we briefly address at the end of Section~\ref{subsec:bbdecay}.

\subsection{Bottomonium Decay}
\label{subsec:bbdecay}

The twin bottomonium states shown in Fig.~\ref{fig:HadronSpectrum} resemble the SM's towers of charmonium/bottomonium states, yet have unfamiliar decay modes.    We relegate most details to Appendix~\ref{app:quarkonium}, but summarize key points here.  Like the $\zpp$, the $\chib$ states can potentially  decay via annihilation to SM 
states through an $s$-channel Higgs.  However, if twin neutrinos (and/or taus) are light,
they can also decay to spin-one $\hat \Upsilon$ states by twin weak interactions, $\chib\to\hat\Upsilon\hat{\ell}\bar{\hat{\ell}}$, where $\ell=\nu,\tau$.  If allowed, this decay competes with the decay via the Higgs, and the competition is very sensitive to modeling of the $\bbtwin$ states.

We can use non-relativistic quantum mechanics to give a rough leading-order estimate of the width of the $\chib$ state to the SM: 
\beq
\Gamma_{\chib \to YY} = 
\frac{27}{4\pi} |R'(0)|^2 \left( \frac{v}{f} \right)^2 \frac{\hat y_b^2}{(m_h^2 - m_\chib^2)^2} \Gamma^{\rm SM}_{h\to YY} (m_\chib^2) \, ,
\eeq
where $R(r)$ is the state's radial wave function.  Here $R'(0)$ appears because $\chib$ is a p-wave state.
As justified in Appendix~\ref{app:quarkonium}, we take the approximation of a 
linear confining potential (with slope $\sigma\approx 4\hL^2$, following calculations of \cite{Juge:1997nd,Morningstar:1999rf}) as a starting point, ignoring the Coulomb potential and important relativistic corrections.  Our estimate is
\beq
\label{eq:chi2SM}
\Gamma_{\chi\to YY} & \sim & 2\times 10^{-3} \left( \frac{v}{f}\right)^4 \frac{m_{\chi}^{11/3} m_0^{10/3}}{v^2 m_h (m_h^2 - m_{\chi}^2)^2} \Gamma_{h \to YY}(m_h)
\eeq
for the lowest $\chib$ state.

 Meanwhile the twin weak decay to $\hat\Upsilon\hat \ell\bar{\hat\ell}$ proceeds via an off-shell $\hat Z$, through a dipole transition (analogous to $\chi_b\to \Upsilon\gamma$ in the SM). Assuming twin neutrinos are massless we roughly estimate
\beq\label{eq:DipoleTransition}
\Gamma_{\chib \to \hat \Upsilon \hat \nu \bar {\hat \nu}} \sim \frac{\hat \alpha_2^2}{4\pi} \frac{(m_\chib - m_{\hat \Upsilon})^7}{(m_{\hat b}
\beta )^2 m_{\hat Z}^4} \, ,
\eeq
where $\beta$ is the typical velocity of the $\btwin$ in this state.  
The extreme dependence of this width on the $\chib-\hat\Upsilon$ mass difference makes any estimate of lifetime and branching fraction highly uncertain.  

Using these formulas as a guide, however, we can qualitatively summarize $\chib$ phenomenology in regions C and D:
\begin{itemize}
  \item If twin neutrinos and taus are not light, then one or more low-lying $\chib$ states may decay promptly at high mass and displaced at low mass.  An approximate formula for the lifetime is 
  given in Eq.~\eqref{eq:ChiToSMRate}.
\item If twin neutrinos and/or taus are light, then
  \begin{itemize}
\item In region C, twin weak decays dominate, making all $\chib$ decays invisible.
\item In region D, the lowest-lying $\chib$ state may decay visibly with a substantial branching fraction, especially for $m_\chib>m_0$.
\item Also in D, the lowest-lying $\chib$ decays may be displaced for $m_\chib>m_0$ and $m_\chib<30\GeV$ or so.
  \end{itemize}
\end{itemize}
 Recall also that in region D there can be mixing of $0^{++}$ glueballs and quarkonia.  This effect is probably of greatest important for $m_0>m_\chib$, where $\zpp$ is nested within the $\chib$ tower.  For $m_0\sim m_\chib$ the quarkonia widths tend to be larger, so the mixed states tend to inherit their properties, with lifetimes somewhat shorter than in Fig.~\ref{fig:hwidth} and the potential for significant invisible decay fractions.

\subsection{Twin Hadron Production}
\label{subsec:twinhadronproduc}

We now discuss twin hadron production via $h$ decays.\footnote{Twin hadron production via the heavy $\hat h$ is discussed in Section~\ref{subsec:heavyhiggs}.}  
Twin hadronization is complicated and quite different from SM QCD, so it is not possible to make reliable estimates as to what fractions of twin glueballs and bottomonia produced are the golden $0^{++}$ states that we can hope to observe.  But the overall rate for $h\to$ twin hadrons, which occurs through $h\to \gtwin \gtwin$ (if $m_0<m_h/2$) and through $h\to \btwin\bar\btwin$ (if $m_\btwin\lesssim m_h/2$),  can be estimated, at least roughly.

We begin with a perturbative analysis of $h$ decays to the twin sector; non-perturbative effects, which can qualitatively change the story, will be discussed later.
We will refer to the $h \to \gtwin\gtwin$ production of glueballs (as shown in Section~\ref{sec:intro}, Fig.~\ref{fig:makingglueballs}) as 
the ``irreducible process'', since, proceeding via the interaction (\ref{eq:dim6}), it is independent of $m_\btwin$ to a first approximation.  
It is the only relevant process for $m_\btwin>m_h/2$, and also dominates for $\hat y_b\lesssim 0.13 y_b$. The partial width for the irreducible process follows from (\ref{eq:dim6}),
\begin{equation}\label{eq:IrreducibleWidth}
\Gamma(h \to \gtwin \gtwin) \simeq \left( \frac{\hat \alpha_3}{\alpha_3} \frac{v^2}{f^2} \right)^2 \Gamma(h \to g g) \, ,
\end{equation}
leading to a perturbative expectation for the branching ratio of the Higgs to twin glueballs of the order $0.1\%$ for $f = 3v$.

Similarly the decay $h\to \btwin\bar\btwin$ can generate $\bbtwin$ states in regions B, C and D, and also glueball states which may be produced immediately along with the $\bbtwin$ ({\it e.g.} $h\to \chib \zpp$), or, if kinematically allowed, in radiative decays ($\bbtwin \to \bbtwin'$ + glueball) or via $\btwin\bar\btwin$ annihilation ($\bbtwin \to$ glueballs).   In region A, $\bbtwin$ states are inaccessible, but still this process can give non-perturbative enhancement to twin glueball production.

At $f\sim 3v$ the perturbative $h\to\btwin\bar\btwin$ width equals the irreducible width for $\hat y_b\sim 0.13 y_b$ and grows as $\hat y_b^2$, enhancing the irreducible rate by about $\sim 60(\hat y_b/y_b)^2$.  This branching fraction becomes so large that the perturbative $h\to \btwin\bar\btwin$ rate  would be inconsistent with current Higgs measurements 
 for $\hat y_b\gtrsim 1.25 y_b$ (at least for $f\sim 3v$; larger values of $\hat y_b$ are allowed as $f$ increases, as we will discuss further in 
 Section~\ref{subsec:precisionhiggs}) --- unless of course $m_\btwin > m_h/2$, in which case only the irreducible process remains.

In sum, within perturbation theory,
\begin{itemize}
\item The high $m_\btwin$ region --- the upper portion of region A, where only glueballs are kinematically allowed --- manifests only the irreducible process, with Br($h\to$ twin glueballs) $\sim 0.1\%$ for $f\sim 3v$, and decreasing like $f^{-4}$. 
\item The low $m_\btwin$ region, including low-mass portions of regions B, C and D, has a substantial $h\to \btwin\btwin$ rate, with branching fraction increasing to $\sim 10\%$ near $m_\btwin\sim 15$ GeV (for $f\sim 3 v)$.
\item Between these two regions the model is ruled out \emph{if we rely on perturbation theory}. However, as we will immediately see, these 
perturbative considerations are not always applicable.  
\end{itemize}

As in SM QCD, where $e^+e^-\to$ hadrons is strongly modulated by resonances and other non-perturbative effects below a couple of GeV, non-perturbative effects are potentially substantial in $h$ decays when kinematics restricts the number and types of hadrons that can be produced thereby.  We discuss this in some detail in Appendix \ref{app:hadroproducdetails}, but here we just summarize our most important observations.
\begin{itemize}
\item
  At moderate to large $m_0$, higher $0^{++}$ glueball resonances (and gaps between them) can enhance (and suppress) the irreducible rate.  A very conservative estimate is that, in the relevant kinematic regime, the suppression can be no worse than 1/10, and is likely less severe.  Thus the irreducible rate is at least $10^{-4}$, and usually larger, throughout the parameter space.
\item At moderate to large $m_{\btwin}$, similar enhancement and reduction may occur from excited $\chib$ states.  If $m_h$ lies between two narrow $\chib$ resonances, then there can be substantial reduction relative to the perturbative $h\to \btwin\bar\btwin$ prediction.  As we discuss in a preliminary fashion in Appendix~\ref{app:hadroproducdetails}, it appears that wherever we can calculate $\chib$ widths and mass-splittings with order-one confidence --- relatively large $m_\btwin$ and relatively small $m_0$ --- the suppression between resonances is always enough to invalidate the perturbative exclusion.  At larger $m_0$ (for fixed $m_\chib$) the widths tend to be even narrower, the splittings larger, and the suppression somewhat greater, though not calculable with current methods.  Due to this non-perturbative suppression, we are confident this model is not universally excluded when $m_\btwin$ lies near and somewhat below $m_h/2$. In fact we believe, for various reasons, that the model is still allowed in quite far into the lower $m_\btwin$ portion of region A, and probably also survives in an upper portion (as well as the lower portion) of region D, but much more work is needed to put this suspicion on a solid footing.
  
\end{itemize}

In summary, we expect the branching fraction for $h\to$ twin hadrons to be at least $10^{-4}$ and, in most of parameter space, much larger.  Final states may include various combinations of twin glueballs and bottomonium.  Although the rates for production of individual types of hadrons cannot be calculated, we expect a substantial fraction of the twin glueballs (and perhaps bottomonia) to be $\zpp$ and $\chib$ states that can decay, possibly displaced, by mixing with the Higgs. While 
$1.25 m_b(f/v)\lesssim m_\btwin<m_h/2$ is excluded {\it within perturbation theory} by existing measurements of the Higgs, this part of parameter space is likely a patchwork of excluded and allowed regions, whose details we cannot calculate without full non-perturbative information.  In the allowed regions, the twin hadron production rate is always substantially enhanced beyond the irreducible rate.

\begin{figure}[t] 
   \centering
          \includegraphics[width=6in]{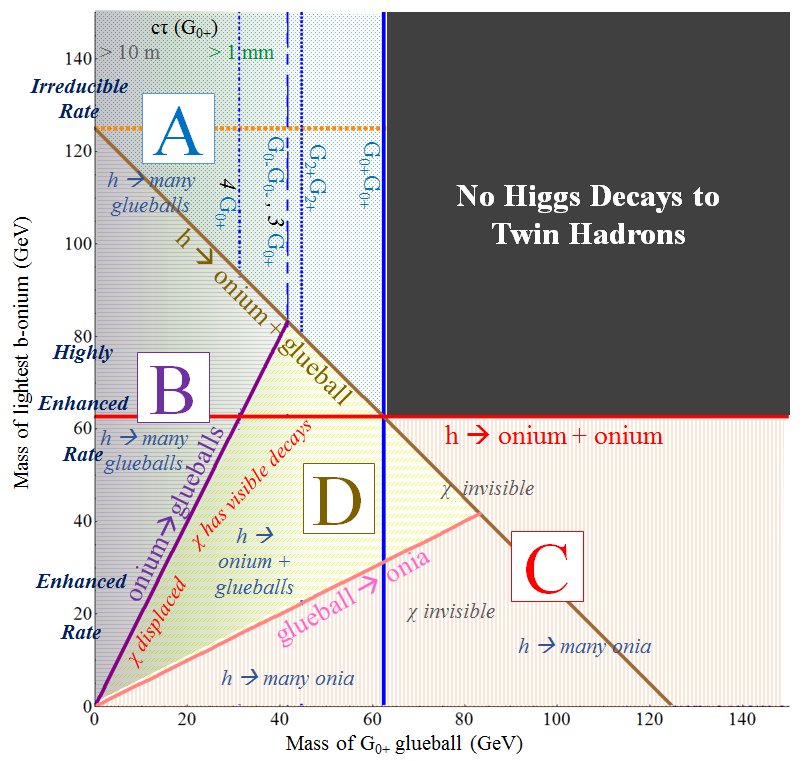}
          \caption{A qualitative overview of the phenomenology, for $f=3v$, in the various regions of parameter space; see Fig.~\ref{fig:Regions0}.  Details are explained in subsequent sections.         
            Solid lines indicate kinematic boundaries.
            Common final states are indicated in italics.  At low glueball mass, decays of the $\zpp$ are displaced; see Fig.~\ref{fig:hwidth}. 
            Here it is assumed that there are light twin leptons, so one $\chib$ state is visible, and even displaced, only in small regions; otherwise $\chib$ decays visibly throughout regions C and D, and is displaced at low mass.}
   \label{fig:Regions}
\end{figure}

A qualitative summary of the phenomenology in the various regions is given in Fig.~\ref{fig:Regions}.  We have conservatively assumed that twin weak decays of 
bottomonium are kinematically allowed, making decays in region C invisible, and making $\chib$ visible only in some limited portion of region D; if instead the twin weak decays are forbidden, $\chib$ decays are visible throughout C and D and displaced at low mass (see Appendix~\ref{app:quarkonium}.)  In the next section we explore these and other signals more carefully.

\section{LHC Phenomenology}
\label{sec:LHCpheno}

By far the most spectacular signal that can arise from our minimal Twin Higgs model is the displaced decays of twin glueballs and quarkonia.   We describe the phenomenology of this signal, as it arises from $h$ decays, in Section~\ref{subsec:zppdecays}; further details on search methods are given in Appendix~\ref{app:strategies}.  If no displaced decays are observable, $h$ decay signals may be challenging for LHC but certainly accessible at a future $e^+e^-$ collider (Section~\ref{subsec:promptdecays}).  A brief discussion of how a twin hypercharge $U(1)$ would affect the phenomenology
is given in Appendix~\ref{app:hypercharge}. Section~\ref{subsec:heavyhiggs} covers signals from  a heavier Higgs $\hat{h}$.
Finally, Section~\ref{subsec:precisionhiggs}, 
explores the order-$(v/f)^2$ effects on SM Higgs production rates.

\subsection{New Higgs Decays With Displaced Vertices}
\label{subsec:zppdecays}

The branching fraction Br($h\to$ twin hadrons) $> 10^{-4}$ everywhere that it is not kinematically forbidden.  Because the number of Higgs bosons produced at LHC in Run II will be of order $10^{7}$, and because displaced 
vertices are spectacular signals when identified, these numbers represent a very promising opportunity.    Already hundreds or thousands 
of events with displaced vertices may have been produced, though in many parts of parameter space they would clearly  have evaded existing 
LHC Run I searches \cite{ATLAShvMS1,ATLAShvMS2,LHCbLLPHiggs,CMSLLPjets,ATLAShvHCAL}.

As Fig.~\ref{fig:Regions} suggests, the model exhibits a great diversity of displaced vertex phenomenology.  Rather than address the full story here, we mainly discuss the regions with the simplest phenomenology.  These regions, it turns out, produce most of the possible Higgs decay signatures, and are therefore sufficient to motivate the most important searches, which are sensitive to effects in more complicated regions that we will not discuss in detail.

The simplest region is the portion of region A with $m_\btwin > m_h/2$, where the irreducible rate applies and only glueballs can be produced.  As we move across this region from large to small $m_0$, taking $f\sim 3v$, we find the following twin hadron phenomena:
\begin{itemize}
\item For $m_0\gtrsim 40$ GeV, $h\to \zpp\zpp$ dominates and $\zpp$ decays are prompt.
\item For $10 \GeV\lesssim m_0\lesssim 40 \GeV$, the $\zpp$ decays are displaced; decays to other glueballs,
  and to higher multiplicities of glueballs, become more common for smaller $m_0$.  The decay $h\to\zpp G'_{0+}$ is of particular note, since the $G'_{0+}$ decays visibly.
\item Below about 10 GeV the $\zpp$ lifetime is so large that decays in the detector are rare. This is partly compensated by higher glueball multiplicity per event.
\end{itemize}

Consider next small $m_\btwin \lesssim 15$ GeV. In region B, where $m_0$ is small, the irreducible process  produces the same phenomena as in region A, but  $h\to\btwin\bar\btwin$ enhances the rate for twin hadron production, leading  (via bottomonium decay to glueballs) to $\geq 3$ twin glueballs per event.  In the low-$m_\btwin$ portion of region C, the glueballs instead decay to bottomonium, leading to states that may all be invisible; alternatively (see Section~\ref{subsec:bbdecay}) final states  may include prompt or displaced $\chib$ decays. In region D, where mixing may be important, and where $\chib\to$ SM has a larger branching fraction, an even richer set of final states is possible.  

  With all of these different subregions with different phenomenological details, many of which cannot be calculated, one may rightly worry that experimental coverage of this model, and others like it, will be extremely difficult.    However, it is possible to bring the challenges under some control by focusing on simple search strategies that cover multiple regions.  For the displaced decays, just a few strategies are potentially sufficient.
\begin{enumerate}
\item Search(es) for single vertex production, $h\to \zpp+\dots$, perhaps separated into
  \begin{enumerate}
  \item $h\to\zpp + \met$ where the $\met$ is due to twin hadrons that decay invisibly and/or outside the detector.
  \item $h\to \zpp + jet(s)$ where a promptly decaying twin hadron produces the jet(s).
    \end{enumerate}
This type of search may only be feasible  when requiring the presence of associated objects that may accompany the $h$, such as a lepton or a pair of vector boson fusion (VBF) jets.    

\item  Search for exclusive di-vertex production production: as in $h\to \zpp \zpp$.
 
\item Search for inclusive di-vertex production: $h\to \geq 3$ twin hadrons, of which at least two decay visibly and displaced.
 
  In contrast to exclusive di-vertex production, here the pair of observed twin hadrons generally have invariant mass {\it below} $m_h$, and need  not be back-to-back in the $h$ rest frame.
  \end{enumerate}

To a limited degree, each of the three search strategies has been explored by 
ATLAS~\cite{ATLAS14LLPLJ,ATLASLLPmutracks,ATLASLLPmuLJ2,ATLASLLPmuLJ,ATLAShvHCAL,ATLAShvMS2,ATLAShvMS1}, 
CMS~\cite{CMSLLPjets,CMSLLPmue,CMSLLPleptons2,CMSLLPleptons} and/or LHCb~\cite{LHCbLLPHiggs}.  However, due often to trigger limitations or analysis gaps, even the most sensitive of these searches do not yet put significant bounds on this model, and a broader and deeper program of searches is needed in Run II.  We will discuss these issues further in Appendix~\ref{app:strategies}.

We note also that although our minimal model has specific relationships between masses, lifetimes and production mechanisms, these relationships will not necessarily hold in other Twin Higgs and Twin Higgs-like models.  It is therefore preferable that the above searches be carried out with the masses and lifetimes of the long-lived particles, and characteristics of the $\met$ (if any), treated as free parameters.  In Appendix~\ref{app:strategies} we suggest benchmark models for these searches and consider some important triggering and analysis issues.

\subsection{New Higgs Decays Without Displaced Vertices}
\label{subsec:promptdecays}

 When the $\zpp$ is heavy, its decays are prompt.  Prompt non-SM decays of the Higgs such as  $h\to \zpp\zpp \to (b\bar b) (b\bar b), (b\bar b)(\mu^+\mu^-)$, or
$h\to \zpp + \met \to (b\bar b)+\met, (\tau^+\tau^-)+\met, (\mu^+\mu^-)+\met$, etc., were examined in a recent 
overview of non-SM Higgs decays~\cite{Curtin:2013fra}.  Not all cases have yet been investigated for a 13--14 TeV machine, but it appears that a non-SM branching fraction of order $10^{-3}$ is far too small for 
discovery of such modes at LHC, though perhaps an $e^+e^-$ machine producing at least $10^5$ Higgs bosons 
could find it.  However, enhancement of the twin hadron branching fraction to $\sim 10\%$, through $h\to \btwin\bar\btwin$ or non-perturbative effects, as discussed in Section~\ref{subsec:twinhadronproduc}, could bring these processes within reach of the LHC.

At the other extreme, one can discuss completely invisible decays.  These can dominate in region C, and can become important in regions A and B if the $\zpp$ has an extremely long lifetime, or in region  D if the $\zpp$ mixes substantially with invisible $\chib$ states.  Other twists on the model (such as the presence of a massless twin hypercharge boson with small kinetic mixing, see Appendix~\ref{app:hypercharge}) can cause all twin hadrons to decay to invisible hidden particles.   Detecting an invisible decay rate much smaller than 10\% is 
very difficult at the LHC, and thus can only be done if there is  significant enhancement by the $h\to\btwin\bar\btwin$ process.
Again, an $e^+e^-$ collider would do much better.

\subsection{Heavy Higgs Decays}
\label{subsec:heavyhiggs}

Not only the $h$ but also the heavy twin Higgs $\hat h$ may serve as a portal, if the twin $SU(2)$  is linearly realized.
The existence of a second perturbative Higgs, while not guaranteed, is favored by precision electroweak data \cite{Espinosa:2012im}.  If present it provides additional opportunities to uncover the twin mechanism.  The mass of this second Higgs is $\sim \sqrt{2 \lambda} f$, and so typically lies around the TeV scale. It possesses a $v/f$-suppressed coupling to top quarks and thus is produced through gluon fusion, albeit with a $(v/f)^2$-suppressed  cross-section compared to an SM Higgs of equivalent mass --- falling at $\sqrt{s}=14$ TeV from 1000 fb at a mass of 300~GeV to 10 fb for a mass of 1~TeV.

On the lower end of this mass range, decays into $WW, ZZ$, and $hh$ dominate, with  branching ratios roughly proportional to $2:1:1$. Once $\hat h \to \hat W \hat W, \hat Z \hat Z$ is kinematically allowed, these processes become at most comparable to $WW, ZZ$, and $hh$; although couplings to the SM bosons are suppressed by $v/f$ due to mixing, the longitudinal coupling scales as $m_{\hat h}^3/v^2$ and entirely compensates.  This is inevitable, since in the Twin Higgs mechanism the $h$ and longitudinal $W,Z,\hat W, \hat Z$ are all Goldstone modes under the same symmetry breaking.

The decays $\hat{h}\to hh, ZZ$ are common in many BSM models, and searches for these promising signals are already underway~\cite{CMS:2014ipa, Aad:2014yja, Chatrchyan:2013mxa, Aad:2013wqa}. The $\hat{h}$ will appear as a resonance with width suppressed by $(v/f)^2$ compared to a SM Higgs of the same mass, times $\frac43 \ (\frac73)$ to account for the channel $\hat{h}\to hh$  (and channels $\hat{h}\to \hat W\hat W, \hat Z\hat Z$, if kinematically allowed.)  Observation of this resonance at equal rates in $\hat{h}\to ZZ$ and $\hat{h}\to hh$, and measurement of its width, could therefore allow for a test of the model.  In Run II,  CMS expects to exclude $\sigma \times \Br (pp \to \hat{h} \to ZZ) \gtrsim 10 \, (4)$ fb for a heavy Higgs of 500 GeV (1 TeV) with 3000 fb$^{-1}$ at $\sqrt{s} = 14$ TeV, or to discover $\sigma \times \Br (pp \to \hat{h} \to ZZ) \gtrsim 30 \,(10)$ fb at $5 \sigma$  \cite{CMS-PAS-FTR-13-024}. This corresponds to an exclusion reach of 900 GeV or a discovery reach of 750 GeV for a $\hat h$ with $f = 3v$.

The $\hat{h}$ may also give rise to twin hadrons, either via $\hat h\to \hat t\bar{\hat t}$ (followed by $\hat t \to \hat b\tau\nu$) or $\hat h \to \hat Z \hat Z$ (with $\hat Z \to \btwin\bar\btwin$). Although the rate is less than the perturbative irreducible rate for $h\to$ twin hadrons, it can easily happen that the trigger efficiency for $h$ decays is low while that for $\hat{h}$ decays is high, so that $\hat{h}$ decays may provide an easier signal.  Moreover the rate to produce twin hadrons via $\hat Z\to \btwin\bar\btwin$ is larger by a factor of 10 or more than $\hat h\to ZZ\to \ell^+\ell^-\ell^+\ell^-$ and $\hat h\to hh\to b\bar b\gamma \gamma$, the cleanest $\hat h\to$ SM processes.  Decays of $\hat{h}$ may often produce displaced glueballs, including ones too heavy to appear in $h$ decays and decaying via an off-shell 
$h$~\cite{Juknevich:2009ji,Juknevich:2009gg}.  Even if all twin hadron decays are prompt, the events might be observable at LHC if the rate, multiplicity, and total (or missing) energy are sufficient~\cite{Strassler:2008fv,swarmofbs}.

On the other hand, if the twin $SU(2)$ is non-linearly realized and the $\hat{h}$ is as indistinct as the $\sigma$ of QCD, then twin hadrons can be produced,  albeit with small LHC rates,  through enhancements of  $gg\to\hat Z\hat Z$ and VBF production of $\hat Z$ pairs.

\subsection{Precision Higgs Measurements}
\label{subsec:precisionhiggs}

\begin{figure}[t] 
   \centering
   \includegraphics[width=3in]{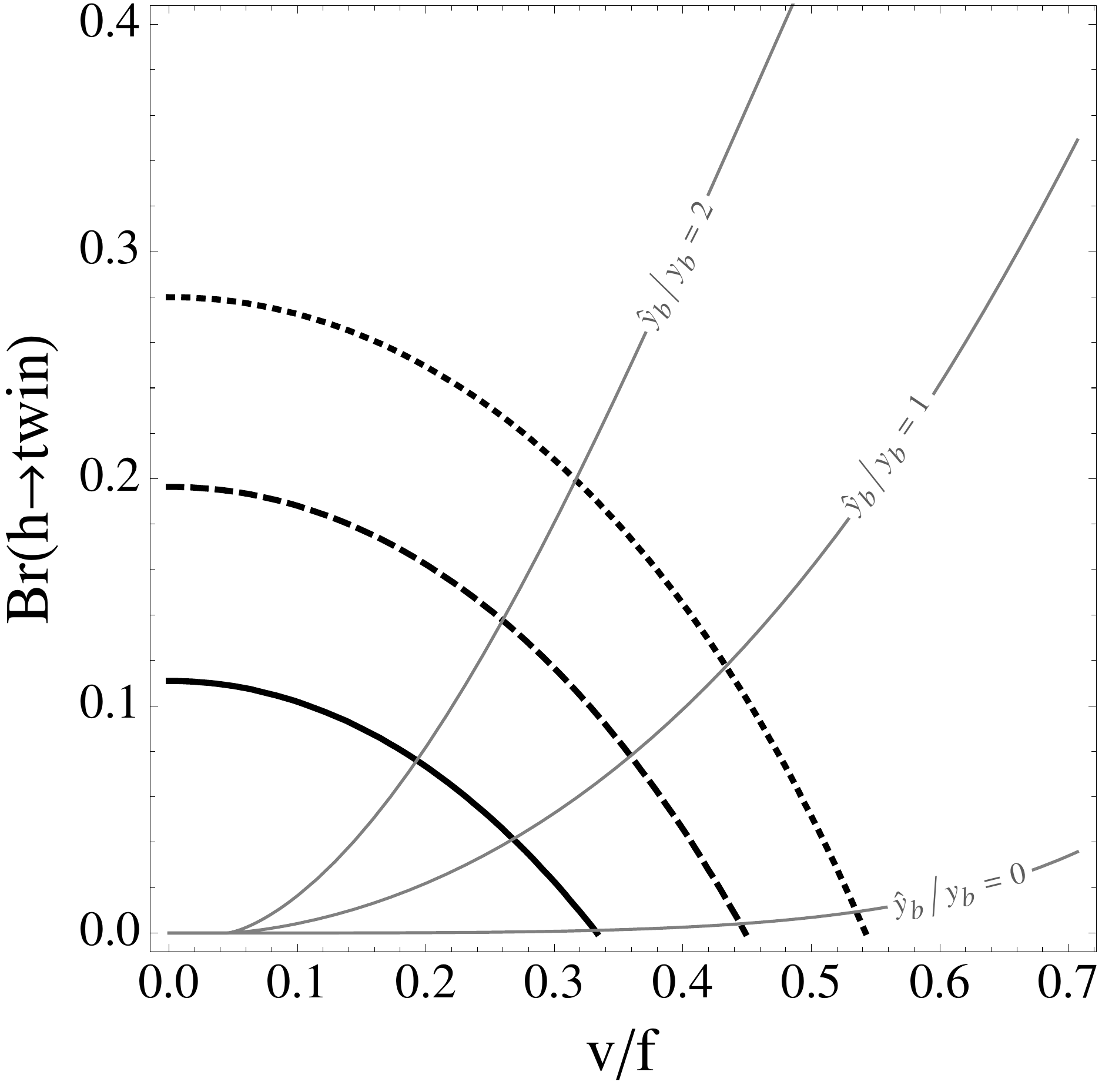} 
   \caption{Current bounds on $v/f$ and the Higgs branching ratio into the twin sector from a combined fit to Higgs coupling measurements. Solid, dashed, and dotted black lines denote the 1-, 2-, and 3-$\sigma$ bounds (defined as $\Delta \chi^2 = 2.30, 6.18, 11.83$) due to ATLAS and CMS Higgs coupling measurements. The grey lines correspond to the perturbative calculation of the Higgs branching ratio into the twin sector as a function of $v/f$ for $\hat{y}_b /y_b = 0,1,2$; as discussed in the text, the actual branching ratio may differ significantly from the perturbative result for a given value of $\hat{y}_b$. }
   \label{fig:fit}
\end{figure}

Here we consider the role of the canonical signature of the Twin Higgs~\cite{Chacko:2005pe}, namely $\mathcal{O}(v^2/f^2)$ changes in Higgs couplings due to the misalignment between the electroweak vacuum expectation value and the pseudo-Goldstone Higgs. This leads to a suppression of all Higgs couplings by an amount 
$1 - \frac{v^2}{2 f^2}$ relative to Standard Model predictions. There may also be a shift in branching ratios due to the additional partial width of decays into the twin sector, but as we have seen this can be much less than 10\%.  Assuming that Br$(h\to$ twin sector) $\ll 10\%$, then the sole effect of the twin sector on SM Higgs measurements is a reduction in 
all production rates, relative to SM predictions, by a factor of $1- \frac{v^2}{f^2}$. To evaluate the impact of current 
Higgs coupling measurements on $v/f$, we have performed a combined fit of the most recent ATLAS and CMS Higgs 
measurements~\cite{Aad:2014xzb, ATLAS-CONF-2014-011, ATLAS-CONF-2014-061, ATLAS-CONF-2014-060, Aad:2014eva,Aad:2014eha, CMS-PAS-HIG-14-009, CMS-PAS-HIG-14-010,  Khachatryan:2014ira} using the profile likelihood method \cite{Cowan:2010js}. The resulting bounds on $v/f$ are shown in Fig.~\ref{fig:fit} as a function of $v/f$ and the Higgs branching ratio into the twin sector.\footnote{This fit does not include implicit precision electroweak bounds from infrared contributions to $S$ and $T$. However, as we will discuss more in Appendix \ref{app:precisionEW}, in contrast to composite Higgs models where the infrared contribution is cut off by $m_\rho \sim $ few TeV and provides the strongest constraint on coupling deviations  \cite{Espinosa:2012im}, here the infrared contribution is cut off by the mass of the heavy Higgs. For $m_{\hat{h}} \lesssim $ TeV these corrections to $S$ and $T$ are comfortably compatible with current precision electroweak bounds and do not strongly influence the coupling fit.} We also show contours corresponding to the perturbative calculation of ${\rm Br}(h \to {\rm twin\; sector})$ as a function of $v/f$ for $\hat{y}_b /y_b = 0,1,2$.  As discussed in Appendix~\ref{app:hadroproducdetails}, the complications of bottomonium production suggest that the actual branching ratio is potentially much smaller than the perturbative value 
 for sufficiently large $\hat{y}_b/y_b$, 
while the irreducible rate for glueball production applies  for $\hat{y}_b /y_b = 0$. 
Current measurements of Higgs couplings place a bound on $v/f$ consistent with the benchmark value $f = 3 v$ considered here. 

For $f \sim 3v$ the shift may be detected definitively before the end of the LHC, but not soon --- certainly not within Run II.  Future projections are a somewhat delicate matter, as measurements in certain channels will become systematics-limited and naive combinations neglecting correlated systematics are no longer appropriate. However, the collaborations quote appropriate coupling projections taking these effects into account. For example, ATLAS projects sensitivity to a uniform scaling factor $\kappa$ of 3.2\% (2.5\%) assuming current (no) theory systematics with 300 fb$^{-1}$ \cite{ATL-PHYS-PUB-2013-014}, which corresponds to a 95\% CL bound of $v/f \lesssim 0.35 \ (0.31)$ assuming the invisible branching ratio gives a sub-leading correction to the observed Higgs couplings. The remaining high-luminosity run should improve sensitivity to 2.5\% (1.6\%) assuming current (no) theory systematics with 3000 fb$^{-1}$ \cite{ATL-PHYS-PUB-2013-014}, leading to bounds of $v/f \lesssim 0.31 \ (0.25).$ It is therefore unlikely that Higgs coupling measurements at the LHC could be used to substantially constrain the parameter space of Twin Higgs models, though if the twin mechanism is operative it may lead to $1-2\sigma$ deviations by the end of the high-luminosity run. (Direct limits on invisible decays are likely to fare even worse, with realistic projections suggesting 95\% CL bounds will approach 8\% with 3000 fb$^{-1}$ \cite{ATL-PHYS-PUB-2013-014},  well short of the typical rate  for decays into the hidden sector.)

 In any case, an overall coupling reduction is purely a sign of mixing and is non-diagnostic; while observation of this reduction would be revolutionary, no unique interpretation could be assigned to it.  Only discovery of the twin hadrons and/or discovery and study of the mostly-twin heavy Higgs would allow for a clear interpretation of such a measurement.

\section{Conclusions}
\label{sec:conclu}

The ``Twin Higgs'' mechanism provides an existence proof for the unsettling possibility that the solution to the hierarchy problem involves a sector of particles that carry no Standard Model quantum numbers, and are therefore difficult to produce at the LHC.  The existence of dark matter already motivates us to consider hidden sectors, and  it is important that the possibility of hidden naturalness be thoroughly considered.

Fortunately, hidden sectors are often not as hidden as they first appear.  As we have seen, a Fraternal Twin Higgs model, one whose hidden sector is not a precise twin of the Standard Model, but which contains the minimal ingredients for the theory to address the hierarchy, naturally leads to Hidden Valley phenomenology.  Specifically, the Higgs sector requires a twin top, which in turn favors twin color; and the lack of light twin quarks in this minimal model then leads to confinement and twin hadrons that include twin glueballs, the lightest of which necessarily mixes with the Standard Model-like Higgs, and twin quarkonium, whose tower of $0^{++}$ states has analogous mixing with the Higgs.  We have shown that the resulting mixing can often cause these glueballs, and  in some cases the lightest quarkonium states, to decay on an observable timescale, leading to a new source of visible non-Standard-Model Higgs decays. The branching fraction for these decays is at worst $10^{-4}$, typically at least $10^{-3}$, and possibly as large as allowed by Higgs measurements.  For heavier twin hadrons with prompt decays, the visible final states --- four heavy-flavor fermions, or two heavy-flavor fermions plus missing energy --- are among those summarized in a recent overview of non-SM Higgs decays~\cite{Curtin:2013fra}. Searches for these final states and for invisible decays at LHC are possible if the branching fractions approach 10\%; if much smaller, a lepton collider may be needed.   But for  moderately light glueballs (and perhaps quarkonia),  these decays can produce a potentially spectacular signal of one or more highly displaced vertices, often accompanied by moderate amounts of missing energy from other twin hadrons that escape the detector.   While several experimental searches for similar vertices have already taken place, a wider array of more powerful search strategies is required if the parameter space for this model is to be fully covered.   

The minimal Twin Higgs model we have presented is only mildly tuned, and no more unnatural than bottom-up modeling  of Higgs compositeness or (natural) supersymmetry. This model also has no obvious cosmological problems, flavor problems, or other glaring issues.  Thus we see no reason from the bottom up, given present knowledge, to treat Twin Higgs as less motivated than, say, composite Higgs models.
Meanwhile, in addition to its cousin Folded Supersymmetry, the Twin Higgs model has recently been generalized, by recognizing it as an orbifold model \cite{Craig:2014aea, Craig:2014roa}.  Variants of these generalized models, by their very construction, share features with our minimal model, though they are different in details.  This shows that the model space of sibling Higgses has not yet been fully explored, and should provide additional motivation for considering seriously and more generally the possibility of a hidden-sector solution to naturalness.  

On general grounds, we expect anything vaguely resembling a Twin Higgs model with a hidden sector to require,  as part of its solution to the naturalness problem, Higgs mixing via a ``portal''-type interaction.   This feature easily leads to additional Higgs-like resonances, new sources of missing energy, and exotic phenomenology of hidden-valley type, including non-SM Higgs decays to multi-body final states and/or displaced vertices.  The challenges are that no individual model is required to produce any or all of these signals, and that production rates for these phenomena are not determined by known interactions (in contrast to gluino or stop production) and can be small.  Among the most motivated places to search for new signals are in decays of known particles, whose production rates are known and large. New decays of the Higgs may not even be rare.  In non-minimal models there may also be opportunities in rare decays of the $Z$.   Searches for new phenomena generated by rarely produced heavy particles (such as the heavy Higgs in our model) must also be considered.

Our field has tended to assume that the solution to the hierarchy problem lies in particles that resemble the ones that we know.  While hidden sectors are often found in string theory vacua and required in models of supersymmetry breaking, their role has been limited to higher energy or purely gravitational interactions, leaving them, as far as the LHC is concerned, out of sight and out of mind.   But with the possible scale of supersymmetry receding upwards, and with no sign of Higgs compositeness or of the colored top partners that were widely expected, the possibility of something more radical, such as a hidden sector around the weak scale that communicates with our sector through a portal, cannot be ignored.  Our searches must move beyond the easier and more obvious lampposts, for the secrets of nature may lie hidden in the dark.

\

\acknowledgments

We thank Christopher Brust, Zackaria Chacko, Patrick Draper, Roni Harnik, Simon Knapen, Pietro Longhi, Colin Morningstar, and Agostino Patella  for useful conversations.
The work of M.J.S. was also supported by NSF grant PHY-1358729 and DOE
grant DE-SC003916.
RS was supported by NSF grant PHY-1315155 and by the Maryland Center for Fundamental Physics.

\appendix{}

\section{Quarkonium Mass Spectrum and Decays}
\label{app:quarkonium}

\subsection{Spectrum of the Quarkonium States}
As explained in Section~\ref{subsec:fratconfinemen}, the most phenomenologically 
interesting quarkonium states are the tower of $0^{++}$ states $\chib$, which can mix with the
SM-like Higgs $h$ and potentially decay back to the SM. 
 We can make a crude but useful analysis of  these states assuming they are governed by non-relativistic quantum mechanics; this approximation holds for  low-lying $\chib$
 states in the regime $m_\btwin \gg \hL$.  The effective quark-antiquark potential can be modeled as a combination of a (logarithmically corrected) Coulomb potential and a long-distance linear potential.
However, $\chib$ decay to SM particles is only of interest if $\chib\to \zpp\zpp$ is kinematically forbidden, since otherwise the twin hadronic decay will dominate.  This requires $m_\chib<2m_0\approx 13.6 \hat \Lambda_3$. One may then check that (remembering a p-wave state is larger than an s-wave state) the Coulomb potential would imply an interquark distance of order $\hat{\Lambda}_3^{-1}$ for any relevant portion of the $(m_0, m_\chib)$ plane, showing the Coulomb approximation is very poor.  
This motivates using a purely linear potential, an approximation which improves for  heavier states. 

Taking the potential 
as linear in $r$ with a string tension $\sigma \approx 4\hL^2$, the Schr\"odinger equation 
for the p-wave radial function is:
\beq
\left[ \frac{1}{2\mu} \left( - \frac{1}{r^2} \frac{d }{dr} r^2 \frac{d}{dr} + \frac{2}{r^2} \right) + 
(\sigma r -E_n) \right] R(r) = 0 \,, 
\eeq
with reduced mass $\mu = \frac{m_\btwin}{2}$. No exact solutions of this equation are known, but an excellent approximation may be obtained by neglecting the linear potential near $r=0$ and the p-wave repulsive potential at large $r$, and then matching the two asymptotic oscillating solutions.  Making this approximation 
we find the energy states
\beq\label{eq:LinearEnergy}
E_n \approx 2 \left( \frac{3\pi}{2} \right)^{2/3} \frac{\Lambda^{4/3}}{\mu^{1/3}} \left( n + \frac{1}{4} \right)^{2/3}
\eeq
and associated wave functions.  This approximate solution is quite close to the exact numerical solution and may be used as a basis for more accurate estimates, though we have not done so here. In any case we believe that this formula 
correctly captures the parametric behavior of the energy levels and masses.

In the Standard Model, highly excited bottomonium states above 10.56~GeV promptly decay to a pair of $B$ mesons; a similar story applies for charmonium. 
But in this model there are no light twin quarks, and correspondingly no twin $B$ mesons, so the twin bottomonium states remain narrow until their masses are above $2 m_{0}$ or $2 m_\hb + m_0$. In most parts of parameter space these are 
very heavy scales, so that the towers of narrow states extend to much higher $n$ than in the SM.

For the physically interesting values of $m_0$ and $m_\btwin$, our approximation does not survive to high $n$, because relativistic effects become large at rather small $n$.  Still, at small $n$ 
our approximation should work for $m_\hb \gg \hL$, including region~A, 
most of the region~B, and some portion 
of regions~D and~C.
Full study of the spectrum with Coulombic and relativistic corrections  is beyond our
scope, but we expect that many of the results that we describe here will be parametrically valid (with order-one corrections) in the relativistic 
regime as well.

\subsection{Decays}

Here we explore decays of the light $\chib$ states  to the SM via mixing with the Higgs, and to a spin-one $\hat \Upsilon$ state plus twin leptons $\hat \ell=\hat \tau,\hat \nu$ via a radiated off-shell $\hat Z$.

In models where the twin leptons are too heavy to allow $\chib\to\hat\Upsilon \hat\ell\bar{\hat\ell}$,
the lowest-lying $\chib$ state will dominantly decay to the SM, with a lifetime obtained from Eq.~\ref{eq:chi2SM}:
\beq
\label{eq:ChiToSMRate}
 c\tau &\sim &  3\ {\rm mm} \times \left(\frac{30 \GeV}{m_\chib}\right)^{7}
\left(\frac{m_\chib}{2 m_0}\right)^{10/3}\left(\frac{750\ \GeV}{f}\right)^4
\eeq
for $m_\chib>10$ GeV. Just as for glueball states (see Fig.~\ref{fig:hwidth}), the lifetime jumps upward once $m_\chib$ falls below the SM $b\bar b$ threshold.  This formula will also apply for the $n^{th}$ $\chib$  state (with an additional $n^{-2/3}$ suppression) if its twin strong and twin weak decays are kinematically forbidden.  In this case, all the lowest-lying $\chib$ states decay visibly even in region C, and displaced decays occur in low-$m_\btwin$ portions of both regions C and D.

Now suppose the twin leptons are light enough that $\chib\to\hat\Upsilon \hat\ell\bar{\hat\ell}$ is kinematically allowed.  (Note the  $\hat \Upsilon $ states will then always promptly annihilate into twin lepton pairs,  which are invisible in this model, via an off-shell $\hat Z$.)  We saw in~(\ref{eq:DipoleTransition}) that the decay $\chib\to\hat\Upsilon\hat\nu\bar{\hat\nu}$ goes as $(m_\chib-m_{\hat\Upsilon})^7$.  While we could attempt to compute this splitting in the non-relativistic approximation, which would not be unreasonable for the higher $n$ states, we know from SM charmonium that this splitting is strongly affected by spin-related effects that push $m_{\hat\Upsilon}$ up and $m_\chib$ down.  Calculation of this splitting is therefore beyond the scope of our work. 

However, we may use a trick to get a good estimate in a phenomenologically important part of parameter space, namely the part of region D (where both bottomonium and glueballs may be produced, often in the same event) that lies close to region B.  This ``BD-boundary'' lies at  $m_\zpp \approx m_\hb\approx \frac{1}{2}m_\chib$, a mass ratio very similar to the SM charmonium system,  for which lattice studies give a
SM glueball mass of 
$1.5 - 1.7$~GeV~\cite{Bali:1993fb,Chen:2005mg,Ochs:2013gi}, and a charm quark 
mass $m_c \approx 1.5$~GeV.  Therefore
we can take results from charmonium data~\cite{Beringer:1900zz},
\beq
m_{\chi_{c0}} = 3.415~{\rm GeV}, \ \ \ \ m_{J/\psi} = 3.097~{\rm GeV},
\eeq
and rescale all mass scales by a common factor to obtain a realistic value for $m_\chib-m_{\hat\Upsilon}$ near the BD-boundary.

\begin{figure}[t]
\centering
\includegraphics[width=.47\textwidth]{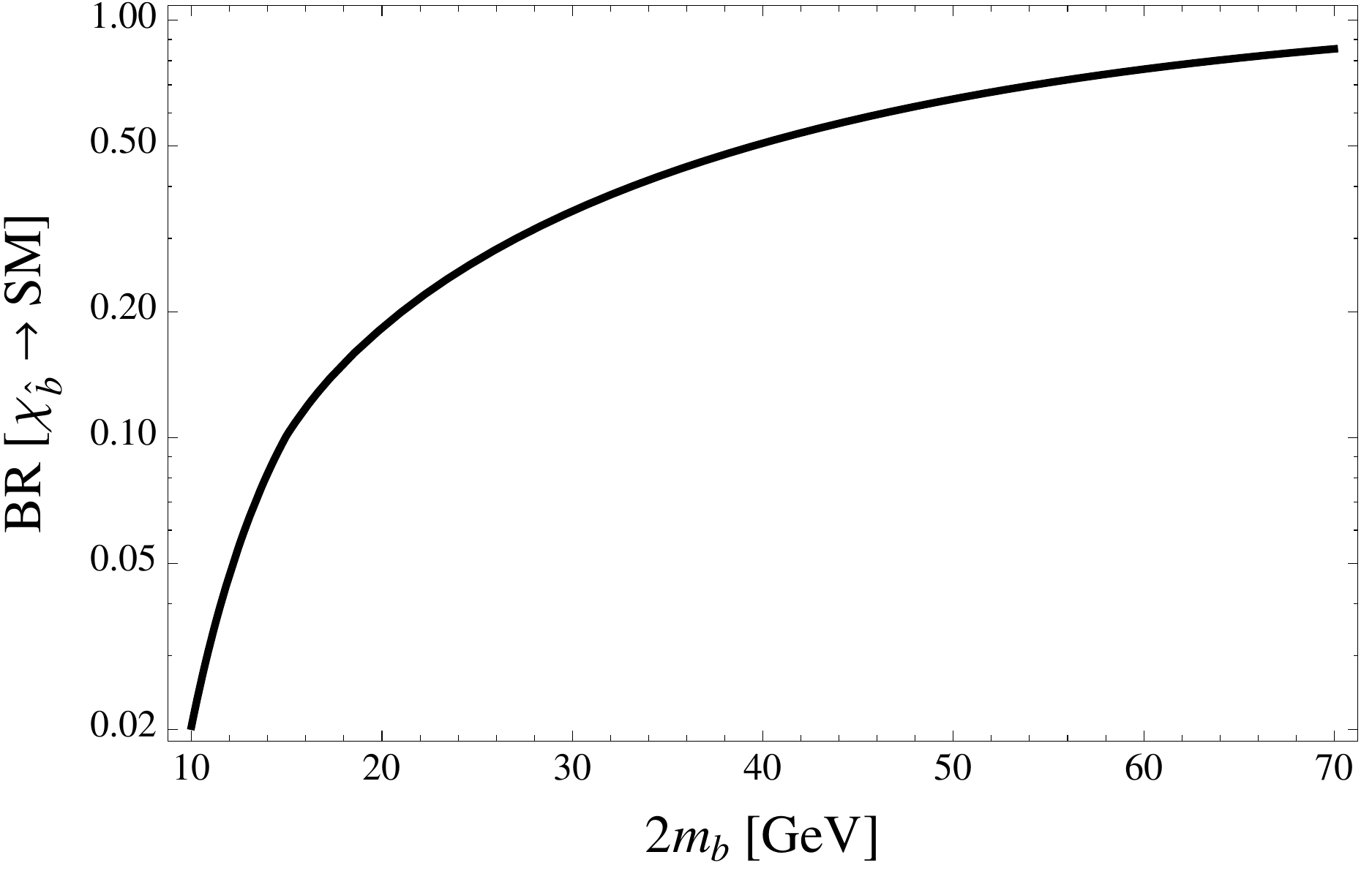}
\includegraphics[width=.47\textwidth]{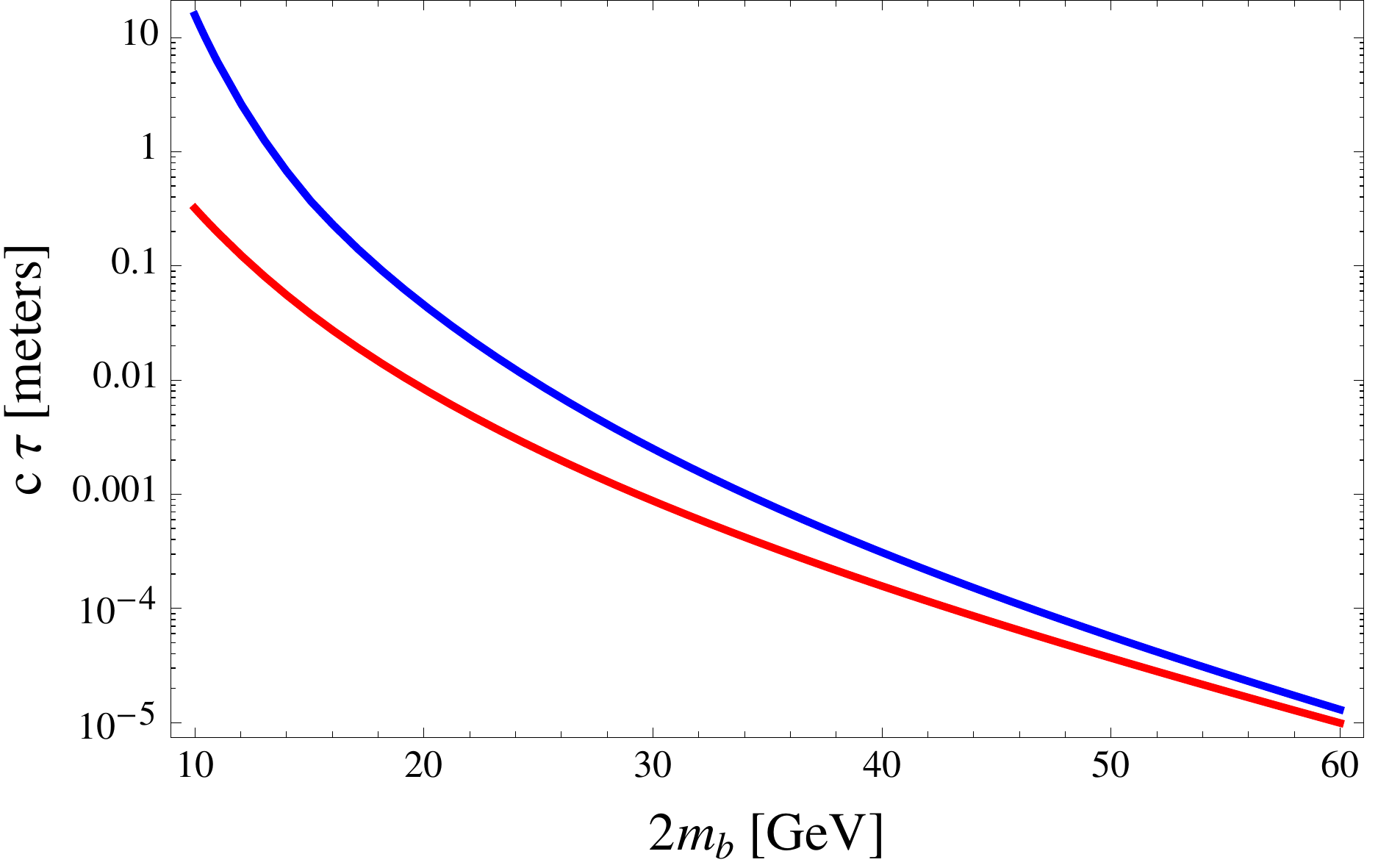}
\caption{Features of $\chib$ decay; we emphasize that these results have very substantial uncertainties due to the limitations of our estimation methods.  Left: Branching ratios of $\chib$ into the SM along the BD-boundary line $m_\zpp = m_\hb$, assuming twin neutrinos are massless.  Right: Decay length of $\chib$ for $m_\zpp = m_\hb$, where the lower red (upper blue) curve refers to massless (heavy) twin neutrinos.}
\label{fig:BDline}
\end{figure}

Using this scaling argument and the formulas of Section~\ref{subsec:bbdecay}, we see from Fig.~\ref{fig:BDline} that displaced decays are possible for $m_\chib$ somewhat below 30 GeV, with lifetimes potentially reaching 10 cm; otherwise the decays are prompt.  The branching ratio of the lightest $\chib$ to SM states is larger than $\sim 10\%$ (larger than $\sim 1\%$) for $2m_\hb \sim 25$~GeV (10~GeV).  We also remind the reader that in this part of region D one expects the Higgs branching rate to twin bottomonium, from the $h\to\btwin\bar\btwin$ process, to be well enhanced above the irreducible rate.  Consequently even a rather small branching fraction can be interesting.

However, we must emphasize that our estimates are highly uncertain due to our crude methods, which include the fact that we have not computed the numerical coefficient in (\ref{eq:DipoleTransition}).  One should therefore only conclude that there is good reason to believe that the lowest $\chib$ may be observable near the BD boundary, and possibly displaced, even if twin leptons are light.  Excited $\chib$ states will decay invisibly.

As we increase $m_0$, moving away from the BD-boundary across region~D and into~C, we cannot reliably compute the $m_0/m_\btwin$ dependence of the $\chib-\hat\Upsilon$ splitting and corresponding twin weak decay rates.  However, both our approximations and reasoning from QCD data suggest the mass splitting rapidly increases with $m_0$, at least as fast as $\hL\propto m_0$ itself.  We therefore expect that even the lowest $\chib$ state decays invisibly with a branching fraction near unity, across the right side of region D and all of region C.

Finally let us note that we are not aware of any other bottomonium states that have observable decays.


\section{Twin Hadron Production in More Detail}
\label{app:hadroproducdetails}

Twin hadrons are produced when $h\to\gtwin\gtwin$ or $\btwin\bar\btwin$, but twin hadronization is complicated and poorly understood.  Despite many years of experimental study of QCD hadron production, theoretical understanding of hadronization is still limited.  Moreover a twin sector with no light twin quarks has a very different spectrum and dynamics from SM QCD.

Perturbation theory applies for inclusive calculations, such as Br($h\to$ twin hadrons), when colored particle production occurs at distances short compared to $\hL^{-1}$, so that $\hat\alpha_s$ is small.  (A similar example is the hadronic branching fraction of the $Z$.)
However, for large $m_0$ or $m_\chib$, twin hadron production is analogous to production of QCD hadrons within or just above the QCD resonance region, where non-perturbative effects can be important.  Decays of the $h$ can be enhanced relative to perturbative estimates if $m_h$ lies close to a narrow excited $0^{++}$ glueball or quarkonium resonance, and suppressed if it lies between two such resonances.

\subsection{Nonperturbative effects in twin gluon production}
\label{subapp:nonperthgg}
First consider $h\to \gtwin\gtwin$ for $m_\btwin>m_h/2$.  For small $m_0$, the glueball resonances near 125~GeV are likely to be both numerous and wide, blending into a continuum.  In this case the irreducible rate is given by the perturbative result~(\ref{eq:IrreducibleWidth}) for $h\to\gtwin\gtwin$.

  For larger $m_0$, resonance effects are potentially important.  Based on large-$N_c$ counting and experience from the $\rho$ meson --- $\Gamma_\rho\sim m_\rho/5$ is of order $1/N_c=1/3$ --- we expect that excited $0^{++}$ glueball resonances have $\Gamma/m\sim 1/N_c^2=1/9$, perhaps with an additional minor suppression factor.  There is also likely a phase-space effect increasing the width for higher excitations, but we ignore this for our conservative estimate. Meanwhile the spacing between $0^{++}$ glueball resonances is likely to be smaller than $m_0$ (as is the case for $\rho$ mesons, among others) but it is surely no larger; lattice evidence for the second $0^{++}$ glueball \cite{Lucini:2004my} supports this.  The maximum suppression between two resonances of width $\Gamma$ and spacing $\Delta$, relative to a perturbative calculation, is of order $\Gamma/\Delta\gtrsim (m_0/N_c^2)/m_0=1/9$.  Thus we expect the worst possible non-perturbative suppression factor is about 0.1, bringing the worst-case rate for glueball production down no further than $10^{-4}$.

  \subsection{Nonperturbative effects in twin bottom production}
\label{subapp:nonperthbb}  
  If the Higgs can decay to $\btwin$ quarks and thus to $\bbtwin$ states, the rate for twin hadron production is often enhanced.  As before, the $\bbtwin$ production rate is given by the perturbative rate for $h\to\btwin\btwin$ for sufficiently small $m_\btwin$ and $m_0$.  As discussed in Section~\ref{subsec:twinhadronproduc}, the perturbative branching fraction for $h\to\bbtwin+X$ grows as $\hat y_b^2$, contradicting existing Higgs measurements for $\hat y_b\gtrsim 1.25 y_b$.

  But at high $\hat y_b$, the perturbative rate often gives the wrong answer.  The widths of the excited $\chib$ states with mass $\sim m_h$ may be very small compared to their mass splittings, due to kinematic constraints.  For instance, the $\chib$ widths are tiny if $m_h < m_{\etab} + m_0, 2 m_\etab$,  and may be quite narrow until $m_h-m_{\etab}\gg m_0$.  If $m_h$ lies between two resonances, then there can be a strong  non-perturbative suppression compared to the perturbative rate.

  We may make an estimate of the maximal suppression factor as we did for glueballs for large $m_\chib$ and small $m_0$. Here the annihilation decay $\chib\to\gtwin\gtwin$ is perturbative and depends on $R'(0)$, which we have estimated as described in Appendix~\ref{app:quarkonium}; this gives us $\Gamma$ if radiative decays of the $\chib$ are sufficiently suppressed by small twin bottomonium mass splittings.  Similarly we may obtain $\Delta$ for the higher $\chib$ states from (\ref{eq:LinearEnergy}).  We find a suppression factor of order
  \beq
  \hat \alpha_3^2 (E_n/m_\chib)^{3/2} \, ,
  \eeq
  where $\hat \alpha_3$ arises from the annihilation rate and should be evaluated at the scale $m_\btwin$, and where $E_n$ is the excitation energy of the relevant $\chib$ state.  Since the relevant $\chib$ state has a mass $\approx m_h$, $E_n\sim m_h-2 m_\btwin$, as long as $n$ is not too small.  This suppression factor can be of order 0.01-0.1.  A suppression factor of $(\hat y_b/1.25)^2\approx (2m_\btwin/30\GeV)^2$ would be enough to make a perturbatively excluded region allowed, so we conclude that some regions with $m_\btwin$ somewhat below $m_h/2$, specifically those where $m_h$ lies between two excited bottomonium states, are probably not excluded by data.

  For larger $m_0$ there can be even more suppression because the rate $\chib\to \gtwin\gtwin$ can be modulated by excited glueballs for the same reason as $h\to \gtwin\gtwin$ is modulated by excited glueball resonances.  We therefore expect that the non-perturbatively allowed portion of region A extends to lower $m_\btwin$ than that of region B.  However, we have no reliable computational methods in this regime.


\section{Search Strategies}
\label{app:strategies}

Here we briefly discuss some triggering and analysis issues with regard to the search for the displaced decays of long-lived twin hadrons.  For brevity, we refer to the visible decay products of a long-lived particle as a ``displaced vertex'' (DV) even when it occurs in regions where tracks are not actually reconstructed.  We also refer collectively to the decay products of a $W$ or $Z$ in $Wh$ and $Zh$ production, and to the vector boson fusion (VBF) jets in VBF Higgs production, as ``associated objects'' (AO).

\subsection{Comments on Triggering}
\label{subapp:triggering}
Triggering on Higgs decays to long-lived neutral particles requires three classes of trigger strategies, which respectively focus on
\begin{itemize}
\item the presence of one or two DVs; this method is largely independent of how $h$ is produced and is sensitive to non-Higgs production of the DVs.  Displaced decay objects can include jets with displaced tracks, trackless jets (possibly including a muon), narrow trackless jets with little electromagnetic calorimeter (ECAL) deposition, and unusual clusters of hits or tracks in the muon system.  
\item the presence of AOs that accompany the Higgs, including VBF jets or daughters of a $W$ or $Z$ (leptons, neutrinos, jets); this method is relatively independent of the details of the $h$ decay and can therefore be used for any exotic $h$ decay mode.
\item the presence of both; for instance, in an event with VBF jets along with a trackless jet.  Requiring both  may be used to lower $p_T$ thresholds on the trigger objects, or to access DVs that would be unusable on their own.  However, this powerful method is specifically optimized for Higgs decays to long-lived particles.
\end{itemize}
Triggers of the first type were used at 
ATLAS~\cite{ATLAStrignote,ATLASLLPTriggers,ATLAShvMS1,ATLAShvMS2,ATLAShvHCAL,ATLASLLPmuLJ,ATLASLLPmuLJ2,ATLASLLPmutracks} and CMS~\cite{CMS-PAS-EXO-11-004,CMSLLPleptons,CMSLLPjets} in Run I; lepton and $\met$ triggers are standard, while a VBF trigger was used in 2012 parked data at CMS~\cite{InvisHVBF}; and no triggers of the third type were used in Run I, to our knowledge.

Depending on the lifetimes of the long-lived particles and on their masses and other kinematics, any one of the three approaches to triggering can work for any of the three search strategies outlined in Section \ref{subsec:zppdecays}.  We leave the appropriate studies to our experimental colleagues.

\subsection{Benchmark Models}

Next we point out possible benchmarks that may be used as straw-man targets.  Although the parameters of the benchmarks are correlated in the Fraternal Twin Higgs --- for instance the mass and lifetime of the $\zpp$ are highly correlated, Eq.~(\ref{eq:GlueballDecayRate}) --- it is important to account for the many uncertainties in twin hadronization, and even more important, to retain model-independence.  Therefore, notwithstanding the particular features of the Fraternal Twin Higgs, it seems best to ignore these correlations and study the benchmark models across their entire parameter space.  

In Section~\ref{subsec:zppdecays} we suggested three possible search strategies, and we now discuss suitable benchmarks appropriate for each of them.  To keep things simple, we discuss them in the context of phenomena that occur in region A.  However, with a little thought the reader may verify that the same benchmarks would be useful for these and other processes that occur in other regions of parameter space.

\subsubsection{Single Displaced Vertex Search}
\label{subapp:onedv}

To look for a single hadronic DV from $h\to \zpp+\dots$ (and similar decays) is challenging, because of large, difficult-to-measure backgrounds. A CMS search \cite{CMSLLPjets} that required only one DV was not able to set limits on decays of a 125 GeV particle.  Relevant ATLAS searches used single DV events to obtain background estimates on double DV events.
However, such a search {\it is} necessary.  When $c\tau_0 \gg 10$ m (as for $m_0\lesssim 10$ GeV in region A), or if hadronization assures that particles making DVs are rare among twin hadrons, the number of DVs rarely exceeds one per event.

To date no search for a DV has exploited the AOs, the VBF jets and/or lepton(s), that sometimes accompany the Higgs.  To obtain background estimates,  it may be enough to measure the rates for events with neither an AO nor a DV, and with either an AO {\it or} a DV; then if the AO and DV are uncorrelated one may predict the rate to have {\it both} an AO {\it and} a DV.

Also, searches for a single DV can demand a second object from the twin hadrons that do not produce a DV.  This object could be $\met$ (relevant for very long-lived $\zpp$) or  prompt jets, possibly $b$-tagged (relevant for $h\to \zpp G'_{+0}$, where $G'_{+0}\to b\bar b$ promptly.) 

In this context we suggest the following benchmark models (in which all $X_i$ of mass $m_i$ decay to SM final states with the same branching fractions as a Higgs of mass $m_i$, unless otherwise noted).  These benchmarks do not cover all the kinematic possibilities but will serve as a useful initial target.
\begin{enumerate}
\item $h\to X_1 X_2$, where $X_1$ has a long lifetime $\tau_1$, and $X_2\to\met$.
  \item $h\to X_1X_2$, where $X_1$ has a long lifetime $\tau_1$ and now $X_2\to$ SM with a prompt decay.
\end{enumerate}
It is important to consider a range of masses for $m_1$ and $m_2$. For the first benchmark, $m_1\neq m_2$ helps  account for the possibility of $h \to$  many glueballs, of which only one gives a DV and the rest are unseen; then $X_2$ represents a system of $\geq 2$ glueballs, with a large invariant mass.  The signal of the second benchmark may arise from decays such as $h\to\zpp G'_{0+}$ or $h\to \zpp\chib$, which generally have $m_1\neq m_2$.

\subsubsection{Exclusive Double Displaced Vertex Search}
\label{subapp:twodv}

In the heart of region A ($m_0\sim 20-40$ GeV), $h\to \zpp\zpp$ often gives two DV, roughly back-to-back and with invariant mass $\sim m_h$. Several searches for this final state have been undertaken at ATLAS \cite{ATLAShvMS1,ATLAShvMS2,ATLAShvHCAL} and at LHCb \cite{LHCbLLPHiggs}.   Also interesting is $h\to\zpp G'_{+0}$; here one has particles of two different lifetimes and masses. 

In this context we suggest benchmark models of the form
\begin{enumerate}
\item $h\to XX$, where $X$ has mass $m_X$ and lifetime $\tau_X$.
  \item $h\to X_1X_2$,  where now the $X_i$ have different masses $m_1<m_2$ and lifetimes $\tau_i$.
\end{enumerate}
In our minimal model it is sufficient to consider $m_1<m_2<2m_1$ since otherwise $X_2\to X_1X_1$ decays occur.  This will not be true in all models however.

\subsubsection{Inclusive Double Displaced Vertex Search}
\label{subapp:twodvincl}

For $m_0\sim 10-25$ GeV, with lifetimes still within the detector, the probability of producing $>2$ glueballs per event is larger, making it more common to have two $\zpp$ DVs along with one or more invisible glueballs.  
One key difference from the exclusive double DV search is that the two DVs may not be well-separated in the lab frame, and may even tend to be found in the same angular region of the detector. Also, the invariant mass of the two (or more) DVs may be well below $m_h$.

A suitable benchmark highlighting this difference would be
\begin{itemize}
\item $h\to X_1X_1X_2$, where the masses are chosen freely (but typically $m_2\geq m_1$), $X_1$ decays as usual with lifetime $\tau_1$, and $X_2\to\met$.  Here the two DVs will have a broad distribution in angular separation and in the momentum that they carry in the $h$ rest frame.
  \end{itemize}

The one serious caveat is the possibility (perhaps remote in the our minimal model, but not necessarily in other models) that the DVs 
are often clustered~\cite{Strassler:2006im,Strassler:2006qa,Strassler:2008fv,ArkaniHamed:2008qp,ATLAS14LLPLJ}.
A special 
benchmark model may be needed if this clustering is sufficiently common that isolation requirements on DV candidates typically fail.  We suggest
\begin{itemize}
\item $h\to X_0X_2$, followed promptly by $X_0\to X_1X_1$,  $X_1$ decays as usual with lifetime $\tau_1$, and $X_2\to\met$.  By choosing $2m_1<m_0<m_2 \ll m_h$, one assures that $X_0$ is relativistic and the two $X_1$ decays are correlated in angle. 
  \end{itemize}


\section{Precision Electroweak}
\label{app:precisionEW}

\begin{figure}[t] 
   \centering
   \includegraphics[width=3in]{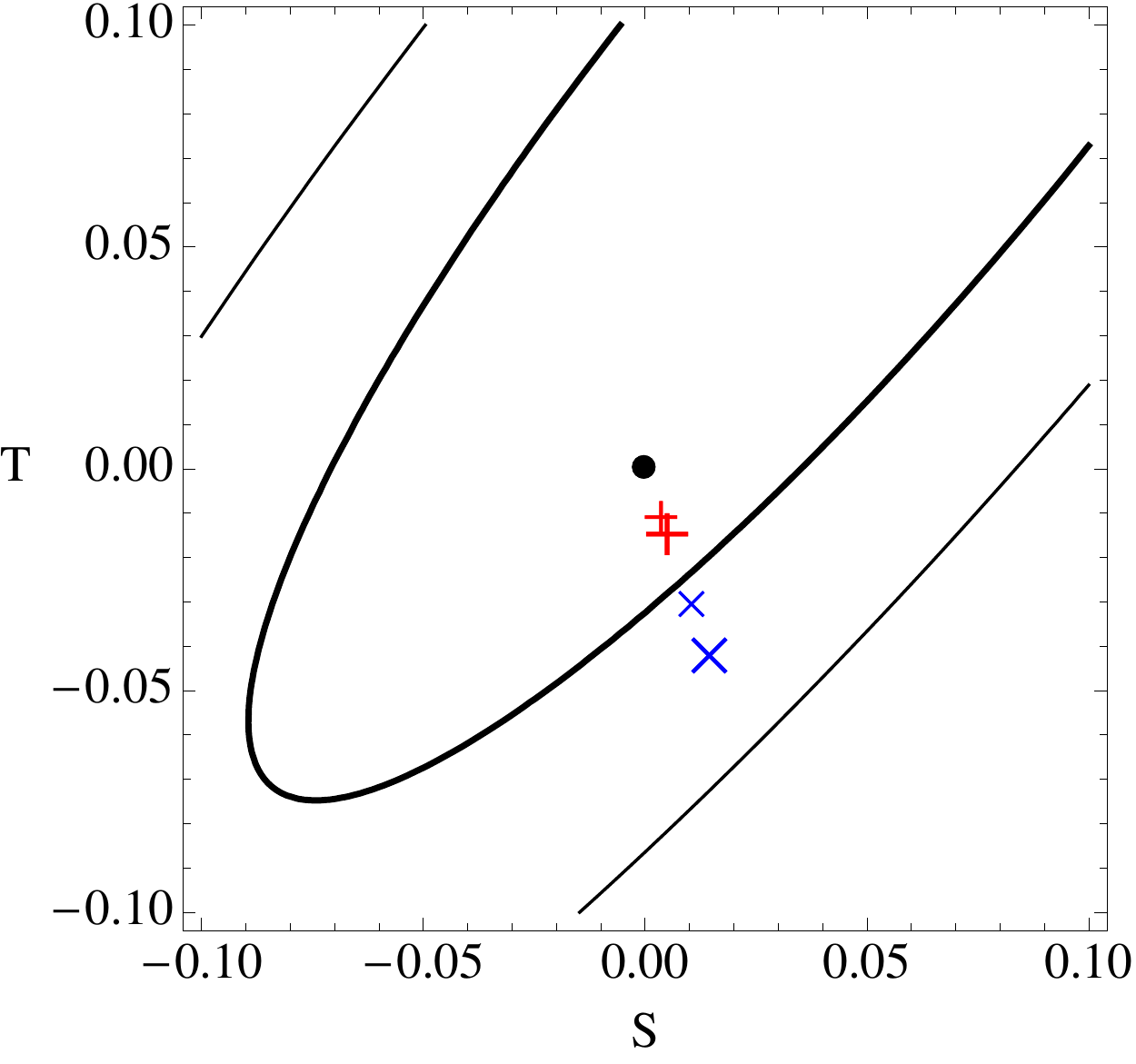} 
   \caption{Close-up of the one- and two-sigma $S,T$ ellipses marginalized over $U$ with contributions from both the SM-like Higgs and the heavy twin Higgs $\hat{h}$. The small (large) blue ``$\times$'' shows the $S,T$ prediction including a heavy twin Higgs of mass $m_{\hat{h}} = 750$ GeV (1.5 TeV) for $f = 3 v$, while the small (large) red ``$+$'' shows the $S,T$ prediction including a heavy twin Higgs of mass $m_{\hat{h}} = 750$ GeV (1.5 TeV) for $f = 5v$. The black dot denotes the contribution of a purely Standard Model Higgs of mass $m_h = 125$ GeV.}
   \label{fig:st}
\end{figure}

Here we briefly consider precision electroweak constraints on the Twin Higgs mechanism. In any theory where coupling deviations of the SM-like Higgs are due to mixing with heavier states there are infrared contributions to the $S$ and $T$ parameters whose coefficients depend on the reduced coupling of the SM-like Higgs to electroweak gauge bosons. The logarithmic contributions to $S$ and $T$ scale like $\log(m_h/m_Z) + (1-a^2) \log(\Lambda/m_h)$, where $a = g_{hVV}/g_{h_{SM} VV}$ and $\Lambda$ is the cutoff scale associated with heavy states. In composite Higgs models the coupling deviations come from mixing with heavy resonances, which must lie in the multi-TeV range due to tree-level contributions to $S$ and $T$. The cutoff on infrared contributions is then of order $\Lambda = 4 \pi v / \sqrt{1-a^2} \sim 4 \pi f$, leading to extremely strong constraints on $a$  \cite{Espinosa:2012im}. 

In the Twin Higgs mechanism, the $\mathcal{O}(v/f)$ deviations in Higgs couplings come from mixing with the heavy twin Higgs and the cutoff is precisely $\Lambda = m_{\hat{h}}$. If the additional twin Higgs is nonperturbatively heavy (i.e.~$m_{\hat{h}} \sim 4 \pi f$), the situation is identical to composite Higgs models, but in general the heavy twin Higgs can be much lighter thanks to a custodial symmetry; there are no tree-level contributions to $S$ and $T$ associated with $\hat h$. The complete contributions to $S$ and $T$ from the SM-like Higgs and heavy Higgs take the form
\begin{equation}
\Delta S \approx \frac{1}{6 \pi} \left( \frac{v}{f} \right)^2 \ln \left( \frac{m_{\hat h}}{m_h} \right) \qquad  \qquad\Delta T \approx - \frac{3}{8 \pi c^2_{\theta_W}} \left( \frac{v}{f} \right)^2 
\ln\left( \frac{m_{\hat h}}{m_h} \right)
\end{equation}
relative to the precision electroweak contributions of a theory with a SM Higgs of mass $m_h$. The current GFitter bound gives $S = 0.05 \pm 0.11$ and $T = 0.09 \pm 0.13$ for a reference Higgs mass of $m_h = 125$ GeV \cite{Baak:2014ora}. We translate this into a fit by marginalizing over $U$ to create the $S,T$ ellipse for free $U$ and plot the spread of IR $S,T$ contributions as a function of $m_{\hat h}$ and $f$ in Fig.~\ref{fig:st}. Remaining within the $1\sigma$ ($2 \sigma$) error for $f=3v$ requires $m_{\hat h} \lesssim 600$ GeV (5 TeV). In contrast to composite Higgs models, precision electroweak constraints do not substantially constrain Higgs coupling deviations in the Twin Higgs as long as the heavy Higgs is in the TeV range.


\section{The Effect of Twin Hypercharge}
\label{app:hypercharge}

In our minimal model, for reasons described in Section~\ref{sec:min}, we have assumed twin hypercharge is absent.  What if it is present?  Then the 
 most 
{\it natural} values of the mass for the twin photon $\hat\gamma$ are (1) $\sim \Lambda$, in which case nothing changes from our present discussion, or (2) zero.

A key issue for twin photons is their kinetic mixing $\frac{\kappa}{2} F_{\mu \nu} \hat{F}^{\mu \nu}$ with the SM photon \cite{Holdom:1985ag}.  In our model the mixing parameter $\kappa$ is not generated at low-loop order at scales below $\Lambda$, so its size is determined by its value at $\Lambda$, which is itself determined by physics in the ultraviolet that we do not specify.  It is thus a free parameter of the model.  

If $m_{\hat\gamma}=0$, we define the twin photon as the massless particle that does not couple to the visible sector, and the $\hat \tau$ and $\hat b$ acquire electric charges $\varepsilon \equiv \kappa \hat e / e$ and $\varepsilon/3$, respectively, in units of $e$. The twin photon is stable and invisible.  All twin hadrons may decay via radiation or annihilation to on-shell or off-shell twin photons --- twin bottomonium through tree-level graphs and twin glueballs through a loop of twin bottom quarks.  The rates for these decays far exceed rates for decays through the off-shell Higgs, and so most decays to the hidden sector are invisible. 

However, any twin photon may be replaced by a SM photon at the cost of $\varepsilon^2 e^2$ in branching fraction.  Constraints on $\varepsilon$ for a massless (or nearly massless) twin photon arise from $e^+e^-$ production of the fractionally charged twin $\hat \tau$ and $\hat b$ fermions.  Since most twin sector production events are invisible, one can obtain a model-independent limit on $\varepsilon$ that depends only on whether twin $\hat \tau$ and $\hat b$ are kinematically accessible, for instance in $Z$ decays.  Such limits are weak, however, of order $\varepsilon \lesssim 10^{-1}$ \cite{Davidson:2000hf}.  But typically the true limits are stronger, because the fraction $\sim\varepsilon^2$ of events that produce a visible signal could have been seen at LEP.  The detailed limits are far too complex for us to work out here.

If $\varepsilon$ is not small, lying in the $10^{-3}-10^{-1}$ range, $h\to\gamma+\met$ and rare $h\to \gamma\gamma+\met$ decays may occur, especially if the twin hadron production rate is large (regions B, C, and D with $m_\btwin$ not too small).  Moreover, annihilation of spin-one states via an off-shell SM photon may produce a visible final state, {\it e.g.} $\hat\Upsilon\to\gamma^*\to \ell^+\ell^-$.  This prompt resonant decay will be subleading by $\sim\varepsilon^2$ compared to the invisible decay $\hat\Upsilon\to \hat\gamma^*\to \hat\tau\hat\tau$ as long as $m_{\hat\tau}<m_\btwin$; otherwise, it competes favorably with highly-suppressed twin sector decays, and may even dominate.  This type of decay is probably common in regions C and D.  In region B, where bottomonium decays to glueballs, this decay can perhaps still occur for the $1^{--}$ glueball, but this glueball is heavy and probably quite rare.

In summary, for $m_{\hat\gamma} = 0$ and at small $\varepsilon$, all twin-sector decays are invisible, but partly-visible decays arise at large $\varepsilon$.  In region A, where production is small, $\gamma+\met$ decays may be the best bet but are very rare. In region B, the larger production rate offers the possibility of a large $\gamma+\met$ rate and a small but important $\gamma\gamma+\met$ rate.  And in regions C and D, the final state $\mu^+\mu^-+\met$, where the $\mu^+\mu^-$ form the lowest $\hat\Upsilon$ resonance, may be either rare or common.  (Higher $\hat\Upsilon$ resonances will often decay radiatively.)  Thus searches for both invisible and partly visible decays of the Higgs are highly motivated; see Section \ref{subsec:promptdecays} above.

Meanwhile, a twin photon with $m_{\hat\gamma}\ll \Lambda$ typically introduces a new naturalness problem if Higgsed, since one more Higgs field $S$ is required and its expectation value and mass must be $\ll \Lambda$. One can get around this if the hypercharge coupling $\hat{g}_1$ is very small, with $\hat{g}_1 \Lambda< m_h/2$. Alternately, the twin photon may be rendered massive by the Stueckelberg mechanism, which is technically natural but entails a parametrically new scale if  $m_{\hat\gamma}\ll \Lambda$. In either case, we view this possibility as lying beyond the {\it minimal} Twin Higgs story.  

For $2m_e<m_{\hat\gamma}<m_h$, the twin photon can play a major role in LHC signals. In contrast to the massless case, here we define the massless SM photon as decoupled from twin particles; the twin photon is identified as the mass eigenstate, which couples to SM fields with strength $\sim \varepsilon e$.  Assuming the twin photon is light enough that all twin hadrons can decay to it, twin hadron production generically (and not just for $0^{++}$ hadrons) leads to high multiplicity ($\geq 4)$ twin photon final states.  Through kinetic mixing with the SM photon, the twin photons then decay to lepton pairs and jet pairs (or simply hadron pairs, such as $K^+K^-$ and $\pi^+\pi^-$, if $m_{\hat \gamma}\sim$ few GeV or below.)  The decays of the twin photon may be prompt or displaced depending on the value of $\varepsilon$, on which there are a number of constraints; for a recent summary, see \cite{Essig:2013lka}.  If prompt, however, limits on multilepton final states (for $m_{\hat\gamma}\gg 1 \GeV$) \cite{Chatrchyan:2011ff,Chatrchyan:2012pka,Aad:2012xsa,ATLAS:2012mn} and on prompt ``lepton-jets''  (for $m_{\hat\gamma}\lesssim 1 \GeV$) \cite{D009LJ,D010LJ,CMS11LJ,CMS12LJ,ATLAS-CONF-2011-076,ATLAS13LJ}
  are strong constraints.   Searches for significantly displaced lepton pairs with masses well above 1 GeV \cite{CMS14LLPLJ} and $\lesssim 1 \GeV$ \cite{ATLASLLPmuLJ,ATLASLLPmuLJ2,ATLAS14LLPLJ} are sensitive to this signal but have by no means excluded it.   Moreover, any such constraints become much weaker once $c\tau_{\hat\gamma}\sim \varepsilon^{-2} m_{\hat\gamma}^{-1} \gg 10$ m. Note also that if some twin hadrons are too light to decay to twin photons, a mixture of these signals and those described in the main text is possible.

\bibliography{fraternal}
\bibliographystyle{JHEP}

\end{document}